\documentclass[screen, acmsmall]{acmart}
\AtBeginDocument{%
  }

\setcopyright{acmlicensed}
\acmDOI{XXXXXXX.XXXXXXX}
\acmConference[Conference acronym 'XX]{Make sure to enter the correct
  conference title from your rights confirmation emai}{June 03--05,
  2018}{Woodstock, NY}
\acmISBN{978-1-4503-XXXX-X/18/06}

\usepackage{graphicx}
\usepackage{listings}
\usepackage{caption}
\usepackage{subcaption}
\usepackage{tabularx}
\usepackage{comment}
\usepackage{balance}
\usepackage{colortbl}
\usepackage{rotating}
\usepackage{mdframed}

\newcommand\new[1]{{\color{black}#1}} 

\newboolean{showcomments}
\setboolean{showcomments}{true}         
\ifthenelse{\boolean{showcomments}}
  {\newcommand{\nb}[2]{
  \fbox{\bfseries\sffamily\scriptsize#1}
     {\sf\small$\blacktriangleright$\textit{\textcolor{red}{#2}}$\blacktriangleleft$}
   }
  }
  {\newcommand{\nb}[2]{}
   
  }
  
\newcommand\masoud[1]{\nb{Masoud}{#1}} 
 
\newcommand\paolo[1]{\nb{Paolo}{#1}}

\newcommand{\COMMENT}[1]{}

\begin{document}
\nocite{*}

\title{A Taxonomy of System-Level Attacks on Deep Learning Models in Autonomous Vehicles}

\author{Masoud Jamshidiyan Tehrani}
\email{masoud.jamshidiyantehrani@usi.ch}
\orcid{0000-0003-3422-4750}
\affiliation{%
  \institution{Universit\`a della Svizzera italiana}
  \city{Lugano}
  \country{Switzerland}
}

\author{Jinhan Kim}
\email{jinhan.kim@usi.ch}
\orcid{0000-0002-0140-7908}
\affiliation{%
  \institution{Universit\`a della Svizzera italiana}
  \city{Lugano}
  \country{Switzerland}
}

\author{Rosmael Zidane Lekeufack Foulefack}
\email{rz.lekeufack@unitn.it}
\orcid{0009-0005-7701-1634}
\affiliation{%
  \institution{University of Trento}
  \city{Trento}
  \country{Italy}
}

\author{Alessandro Marchetto}
\email{alessandro.marchetto@unitn.it}
\orcid{0000-0002-6833-896X}
\affiliation{%
  \institution{University of Trento}
  \city{Trento}
  \country{Italy}
}

\author{Paolo Tonella}
\email{paolo.tonella@usi.ch}
\orcid{0000-0003-3088-0339}
\affiliation{%
  \institution{Universit\`a della Svizzera italiana}
  \city{Lugano}
  \country{Switzerland}
}

\renewcommand{\shortauthors}{Tehrani et al.}

\begin{abstract}
The advent of deep learning and its astonishing performance has enabled its usage in complex systems, including autonomous vehicles. On the other hand, deep learning models are susceptible to mis-predictions when small, adversarial changes are introduced into their input. Such mis-predictions can be triggered in the real world and can result in a failure of the entire system.
In recent years, a growing number of research works have investigated ways to mount attacks against autonomous vehicles that exploit deep learning components. Such attacks are directed toward elements of the environment where these systems operate and their effectiveness is assessed in terms of system-level failures triggered by them. There has been however no systematic attempt to analyze and categorize such attacks.
In this paper, we present the first taxonomy of system-level attacks against autonomous vehicles. We constructed our taxonomy by selecting \new{21} highly relevant papers, then we tagged them with 12 top-level taxonomy categories and several sub-categories. 
The taxonomy allowed us to investigate the attack features, the most attacked components \new{and systems}, the underlying threat models, and the \new{failure} chains from input perturbation to system-level failure. We distilled several lessons for practitioners and identified possible directions for future work for researchers.
\end{abstract}




\maketitle

\section{Introduction}
\label{sec:intro}

The advent of Autonomous Vehicles (AVs) has brought about a paradigm shift in transportation~\cite{kuutti2020survey}. They leverage an array of sensors, cameras, and Deep Learning (DL) models to perceive and navigate their surroundings for safer and more efficient travel. However, as these vehicles have become increasingly reliant on complex DL components, they also expose new attack surfaces. While highly effective in most cases, these DL components possess inherent vulnerabilities, such as being sensitive to adversarial examples~\cite{biggio2018wild}, that can lead to catastrophic failures. A mis-prediction at the DL model level can quickly escalate into a system-level failure, resulting in vehicle crashes or collisions with pedestrians, thus endangering lives. 
\new{Some studies discuss model- and system-level failures under critical conditions, such as bad weather\cite{abdessalem2018testing}. These are considered part of functional testing for AVs. However, they do not evaluate model- or system-level failures under adversarial attacks. Salay et al.~\cite{salay2021missing} pointed out that there is a missing link between system-level safety requirements and component-level performance guarantees, particularly for perception tasks. These tasks are both critical and challenging because they heavily rely on machine learning, which introduces uncertainty and variability in performance. Burton~\cite{burton2022causal} concluded that safety assurance for machine learning requires probabilistic reasoning and a causal understanding of how model insufficiencies can propagate into system-level failures. He also emphasized that relying on a single form of evidence is insufficient to ensure safety.}
A large body of research has focused on the model-level security of DL components, tested by introducing various attack (resp. defense) strategies to induce (resp. prevent) mis-predictions in these models~\cite{cina2023wild}. As evidenced by Wang et al.~\cite{wang2023does}, however, not all attacks on the DL model propagate to the system of an AV, meaning that the success of the model-level attack does not always translate to system failures. In their study, this was largely due to a lack of consideration for the physical properties of the underlying vehicle system, such as maximum and minimum acceleration/deceleration, steering rates, and sensor mounting positions. \new{There are safety standards for autonomous systems, such as ANSI/UL 4600 and ISO/PAS 8800, that highlight the connections between model-level errors and system-level failures. These standards also emphasize the importance of understanding safety-related subsets of the input space and how certain types of errors can contribute to failures at the system level.}
A few recent studies have extended the evaluation of attack methods beyond the model level, considering their system-level impacts in real-world settings, such as the frequency of vehicle crashes~\cite{shen2024soksemanticaisecurity}. 
To the best of our knowledge, no research has comprehensively taxonomized the attacks on AVs' DL models that cause system-level failures.
\new{This paper presents a taxonomy of system-level attacks on AVs, building upon existing research that has explored the results of DL model-level failures under adversarial attacks in the context of AVs. The contribution of this work is to provide an understanding of the vulnerabilities and potential attacks on AV systems, which can help developers identify areas where defenses are most needed.}

We adopted a bottom-up approach to the creation of a taxonomy for classifying the existing literature. We allowed the incremental identification and creation of new categories, as we analyzed the literature in depth. We collected papers by searching scientific databases using a carefully designed query. This initial search yielded \new{9,748 papers}. Then, we manually analyzed and filtered these papers based on their venues, reducing our list to 1,125 potential candidates. Lastly, through snowballing and a thorough assessment by three reviewers, we identified \new{21 papers} that formed the basis of our taxonomy. These papers were categorized into 12 distinct groups, including system-level failure-specific categories such as `System Under Attack' or `System-level Results' to differentiate our taxonomy from previous work that focused only on the characteristics of model-level attacks~\cite{biggio2018wild}.
We further provide statistics to highlight the distribution of properties in the taxonomy, as well as a tree-like visualization to enhance the readability and practical usage of the taxonomy. Finally, our discussion elaborates on the implications and potential future directions arising from our taxonomy and findings, opening up new avenues for future exploration in this emerging field.

The contributions of the paper are as follows:
\begin{itemize}
    \item The first taxonomy of security vulnerabilities that affect autonomous vehicles whose perception components are based on DL. The taxonomy consists of 12 top-level categories and several subcategories.
    \item A detailed analysis of 21 papers that are highly relevant for system-level attacks against autonomous vehicles. We provide a thorough mapping between papers and taxonomy categories.
    \item \new{A comprehensive overview of the AV domain to identify categories covered or missing in the taxonomy papers, highlighting gaps and directions for future research.}
    \item A discussion of the main findings obtained while constructing the taxonomy and analysing the papers. The discussion includes implications for practitioners and possible avenues for future research.
\end{itemize}

The rest of the paper is structured as follows. Section~\ref{sec:background} introduces some important notions about AVs and about DL security. Section~\ref{sec:rqs} comments on the research questions that triggered and motivated the construction of our taxonomy. \new{Section~\ref{sec:domain} analyzes the existing domains of AVs, DL models and simulators}. Section~\ref{sec:methodology} describes the methodology adopted for paper selection and taxonomy construction. Section~\ref{sec:results} presents our findings, including a description of the taxonomy and the mapping of the considered papers. Section~\ref{sec:discussion} answers the initial research questions and discusses the implications of this work for practitioners and researchers. Section~\ref{sec:related_work} comments on the related work. Section~\ref{sec:threats} enumerates the main threats to the validity of our findings and the ways in which we tried to mitigate them. Section~\ref{sec:conclusion} contains our conclusion and anticipates some possible future works.

\section{Background}
\label{sec:background}

\subsection{Autonomous Vehicles}

Vehicles equipped with automated components can achieve different degrees of autonomy. For self-driving cars, five levels have been defined~\cite{J3016_202104}, ranging from \textit{driving assistance} to \textit{full driving automation}. However, according to Faisal et al.~\cite{faisal2019understanding},
an automated vehicle system can only be termed \textit{autonomous} when the vehicle's automated system can seamlessly perform all dynamic driving tasks across a diverse range of environments (i.e., only at level 5, \textit{full driving automation}~\cite{J3016_202104}). In this paper, we are not interested in differentiating among different degrees of automation vs. full autonomy, as we consider in the scope of our taxonomy all works that perform system-level security attacks against the deep learning components used to provide \textit{some} automation to a vehicle, up to full autonomy.

The motion control of a vehicle can be divided into two primary tasks: \textit{lateral} motion control and \textit{longitudinal} motion control. Lateral control, managed by the vehicle's steering, aims to maintain the vehicle's lane position and perform maneuvers such as lane changes and collision avoidance. This is typically achieved through DL components that process images from onboard cameras. Longitudinal control, managed by the gas and brake pedals, focuses on reaching/maintaining the desired speed, ensuring safe distances from other vehicles, and preventing rear-end collisions. This control relies on sensors like radar and lidar, as well as cameras, to measure relative velocity and distance~\cite{kuutti2020survey}. \new{If an AV combines data from different types of sensors to improve accuracy and robustness, it's using a Multi-Sensor Fusion (MSF) system~\cite{li2015survey}. If it merges multiple or single inputs from the same sensor type to enhance perception, it's using a Single-Sensor Fusion (SSF) system~\cite{heng2019project}.
} Other relevant tasks that are assisted by DL components include parking, driver's attention checking, pedestrian and obstacle avoidance, traffic sign/light detection and recognition. \new{Overall, AVs may contain the following modules to achieve  autonomy~\cite{levinson2011towards}:
\begin{itemize}
    \item \textbf{Perception:} Extracts and interprets sensor data to detect objects and understand the environment.
    \item \textbf{Localization:} Determines the vehicle’s precise position and orientation in the world.
    \item \textbf{Planning:} Creates safe and efficient paths based on goals (global planning) and surroundings (local planning).
    \item \textbf{Control:} Executes planned paths by commanding the vehicle’s actuators.
    \item \textbf{End-to-End:} Uses a single model to map sensor inputs directly to control actions.
\end{itemize}
}

\subsection{Adversarial Attacks}

DL models have achieved impressive performance across various application domains. 
However, these technologies can be easily deceived by adversarial examples, such as carefully perturbed input samples designed to mislead detection systems during testing. Researchers have demonstrated that deep neural networks for object recognition or classification can be fooled by input images altered in ways that are imperceptible to humans~\cite{biggio2018wild}. 
A \textit{model-level attack} is an attack that exploits weaknesses of the DL
model in isolation to obtain a mis-prediction for specific, possibly manipulated, inputs.
Specifically, there are two primary \textit{model-level attack} strategies. Firstly, \textit{poisoning attacks} occur when an attacker, with access to the training data, inserts malicious data into the training set. Such data acts as a trigger and when the trigger is present in the input at inference time, it forces the model to make incorrect predictions. Secondly, \textit{evasion attacks} occur when an attacker manipulates input data during testing without altering the model itself~\cite{biggio2018wild,cina2023wild}. These attacks highlight the vulnerabilities of DL models in isolation, but they might be leveraged to trigger potential consequences at the system-level, in safety-critical applications such as AVs. 

\new{To establish the scope of our taxonomy, we first clarify the key concepts used in this work. We define a \textit{model-level attack} as an attack that exploits vulnerabilities and limitations in machine learning models, leading to model-level failures such as misclassification, increased latency, and model uncertainty.} We define a \textit{system-level attack} as an attack that manipulates the environment where the AV operates, in order to cause a system-level failure of the AV. In turn, a \textit{system-level failure} is a deviation of the autonomous system's behavior from its functional/safety requirements. 
When mounting a system-level attack, one or more model-level attacks can be exploited by first \new{forcing} some malicious inputs into the models under attack, and then ensuring the mis-predictions propagate to a system-level failure.
This paper aims to build a taxonomy based on the literature that has explored system-level attacks and their impact in terms of system-level AV failures.

\section{Research Questions} \label{sec:rqs}

Research on system-level attacks against AVs is an emerging, rapidly growing field. 
The \textit{goal} of our work is to acquire a better understanding of the research landscape by categorizing the existing works and defining the main dimensions that characterize them. More specifically, we build a system-level attack taxonomy to answer the following research questions:\\

\noindent
\textbf{RQ1 [Attack Features]:} \textit{What are the prevalent features of the attacks, based on the paper distribution across taxonomy categories?} 

\noindent
With this RQ, we investigate how extensively the research landscape has been explored, to highlight regions that are still relatively unexplored and to identify the features that characterize the most populated regions. This might indicate possible directions for future work, but also the reasons why some research directions are more feasible than others. \\

\noindent
\textbf{RQ2 [Attacked Components]:} \textit{Which components of the system are the targets of the attack?}

\noindent
DL models are used for several tasks in AVs, including lane keeping, obstacle recognition and avoidance, and traffic sign/light recognition, \new{which are components of different AV modules, such as the perception, planning, localization, and control modules.} With this RQ, we investigate if there are preferred attack targets among the DL components commonly integrated into AVs. We aim to identify the features of those components that make them suitable for system-level attacks, possibly operated by manipulating the environment where the AV operates.\\

\noindent
\textbf{RQ3 [Threat Models]:} \textit{What are the threat models?}

\noindent
Different system-level attacks are mounted under very different assumptions about the knowledge available to attackers and their capability to manipulate the environment. Hence, some assumptions may be more reasonable or realistic than others, substantially changing the probability that an attack could be performed in practice. With this RQ, we aim to understand whether the research effort was directed mostly toward realistic attacks that are reasonably possible in the real world, or toward scenarios based on unrealistic assumptions that make them very unlikely. \\

\noindent
\textbf{RQ4 [Consequences]:} \new{\textit{What are the consequences of each attack type at the system level? }}

\noindent
\new{The main difference between a model-level and a system-level attack is that in the latter case, an attack is successful only if model-level mis-predictions influence downstream components and ultimately lead to a system-level failure. With this RQ, we investigate the attack paths exploited in existing works, considering the attack vector (e.g., the environment element manipulated by the attack), its effects on a model input, a model mis-prediction, and the outcomes of such mis-prediction in external observable actions that eventually cause a system failure.}

\section{Analysis of the AV Domain}
\label{sec:domain}
\new{
AVs range widely, from cars and trucks to spacecraft. They rely on various DL models, modular architectures, and simulation environments for development and testing. Before searching for scientific papers that deal with system level attacks against AVs, we compile a comprehensive list of  \textit{AV types}, \textit{DL models}, and \textit{simulation tools} used in this domain.
The goal is to get a thorough understanding of the AV domain that is not biased by the AV types, models and simulators considered in practice in the papers selected for taxonomy construction.
We have utilized a Large Language Model (LLM), specifically the GPT-4-turbo variant~\cite{openai2023chatgpt}, for this task, using a carefully crafted prompt that requires a structured output along with relevant reference links. All outputs and links were manually reviewed to ensure correctness and relevance. We chose to leverage an LLM over traditional web search due to its ability to synthesize domain-specific information from broad, unstructured sources. While LLM-generated content may be prone to hallucinations, our manual verification revealed no instances of hallucinations, misleading, or incomplete information. Based on the responses, we created Table~\ref{tab:AVoverview}, which we overview below. The exact prompt and corresponding outputs are included in the replication package on Zenodo\footnote{https://zenodo.org/records/16407709} and GitHub\footnote{https://github.com/MasoudJTehrani/sys-tax-artifact}.
\begin{table}[]
\caption{An overview of autonomous vehicles, models and simulations}
\label{tab:AVoverview}
\resizebox{\textwidth}{!}{%
\begin{tabular}{cc}
\hline
\multicolumn{2}{|c|}{\textbf{Autonomous Vehicles}}                                                                                                                                                                                                                   \\ \hline
\multicolumn{1}{|c|}{\textbf{Surface}}        & \multicolumn{1}{c|}{\textbf{Types}}                                                                                                                                                                                \\ \hline
\multicolumn{1}{|c|}{Land}                    & \multicolumn{1}{c|}{UGV (including military robots), Autonomous Cars/Trucks/Buses}                                                                                                                                 \\ \hline
\multicolumn{1}{|c|}{Air}                     & \multicolumn{1}{c|}{UAVs/Drones, eVTOL Air‑Taxis, Autonomous Combat/Security Drones}                                                                                                                               \\ \hline
\multicolumn{1}{|c|}{Sea}                     & \multicolumn{1}{c|}{USVs (boats, cargo ships, saildrones, survey vessels)}                                                                                                                                         \\ \hline
\multicolumn{1}{|c|}{Underwater}              & \multicolumn{1}{c|}{AUVs (cylinder, glider, hybrid), Large UUVs, ROVs (remote)}                                                                                                                                    \\ \hline
\multicolumn{1}{|c|}{Space}                   & \multicolumn{1}{c|}{Spacecraft, planetary rovers, autonomous satellites/explorers}                                                                                                                                 \\ \hline
\multicolumn{2}{l}{}                                                                                                                                                                                                                                               \\
\hline
\multicolumn{2}{|c|}{\textbf{Autonomous Models \& Modules}}                                                                                                                                                                                                          \\ \hline
\multicolumn{1}{|c|}{\textbf{Modules}}        & \multicolumn{1}{c|}{\textbf{Common Models \& Techniques}}                                                                                                                                                          \\ \hline
\multicolumn{1}{|c|}{Perception}              & \multicolumn{1}{c|}{\begin{tabular}[c]{@{}c@{}}CNNs (YOLO, Mask R‑CNN, SSD), Panoptic NN (YOLOP), \\ LiDAR networks (VoxelNet, PointRCNN)\end{tabular}}                                                            \\ \hline
\multicolumn{1}{|c|}{Localization \& Mapping} & \multicolumn{1}{c|}{SLAM (EKF, GraphSLAM), Visual odometry with DL-enhanced features}                                                                                                                              \\ \hline
\multicolumn{1}{|c|}{Prediction \& Planning}  & \multicolumn{1}{c|}{LSTM/RNN for trajectories, GNNs (VectorNet), DRL, inverse RL}                                                                                                                                  \\ \hline
\multicolumn{1}{|c|}{Control}                 & \multicolumn{1}{c|}{DRL/Q-networks, inverse RL for policy generation}                                                                                                                                              \\ \hline
\multicolumn{1}{|c|}{End-to-End}              & \multicolumn{1}{c|}{\begin{tabular}[c]{@{}c@{}}CNN “black box” (Openpilot, DAVE‑2), encoder-decoder + depth, \\ MLLM (EMMA), vision-first foundation models\end{tabular}}                                          \\ \hline
\multicolumn{2}{l}{}                                                                                                                                                                                                                                               \\
\hline
\multicolumn{2}{|c|}{\textbf{Simulation Environment}}                                                                                                                                                                                                                \\ \hline
\multicolumn{1}{|c|}{\textbf{Surface Domain}} & \multicolumn{1}{c|}{\textbf{Simulators}}                                                                                                                                                                           \\ \hline
\multicolumn{1}{|c|}{Land}                    & \multicolumn{1}{c|}{\begin{tabular}[c]{@{}c@{}}CARLA, AirSim (cars), Gazebo/Ignition, Webots, CoppeliaSim, \\ Unreal/Unity car sims, SUMO, Drake\end{tabular}}                                                     \\ \hline
\multicolumn{1}{|c|}{Air}                     & \multicolumn{1}{c|}{\begin{tabular}[c]{@{}c@{}}AirSim, Gazebo (PX4 SITL, RotorS), Webots, CoppeliaSim, FlightGear/X-Plane, \\ Pegasus/Isaac Sim, Flightmare, ROSUnitySim, USARSim, MORSE, GrADyS-SIM\end{tabular}} \\ \hline
\multicolumn{1}{|c|}{Sea}                     & \multicolumn{1}{c|}{Gazebo + usv\_sim\_lsa, Marine AUV/USV Simulator (commercial)}                                                                                                                                 \\ \hline
\multicolumn{1}{|c|}{Underwater}              & \multicolumn{1}{c|}{Gazebo + uuv\_simulator, MATLAB \& Simulink AUV Toolbox}                                                                                                                                       \\ \hline
\multicolumn{1}{|c|}{Space}                   & \multicolumn{1}{c|}{Trick (NASA), Gazebo (NASA rover sim)}                                                                                                                                        \\ \hline
\end{tabular}
}
\end{table}

}

\subsection{Typology of AVs}
\new{
AVs encompass a wide array of systems capable of operating independently across various environments, including land, air, sea, and space, by leveraging integrated sensors, artificial intelligence (supervised and reinforcement learning), control systems, and general software. According to the LLM's output, these vehicles can be classified based on their operational domains. Land-based autonomous systems include self-driving cars, trucks, buses, agricultural machinery, and ground robots~\cite{wikipediaUGV, wikipediaSDC}. Airborne vehicles encompass fixed-wing drones, rotary-wing UAVs, vertical take-off and landing (VTOL) aircrafts, and high-altitude pseudo-satellites (HAPS)~\cite{wikipediaUAV}. Surface-sea vehicles (USVs) operate above water for tasks such as marine surveillance and cargo transport~\cite{wikipediaUSV}, while underwater vehicles (AUVs/UUVs) conduct deep-sea exploration and inspection missions~\cite{wikipediaAUV}. Space-based AVs range from planetary rovers to autonomous spacecraft performing orbit control, docking, and planetary landing~\cite{wikipediaVA}.}

\subsection{DL Models in AV Systems}
\new{
AVs rely on a suite of DL models. According to the LLM's output, such models are modularly deployed across perception, localization, prediction, planning, and control systems. In the perception module, convolutional neural networks (CNNs) such as YOLO~\cite{cao2023you}, Mask R-CNN~\cite{he2017mask}, Single Shot MultiBox Detector (SSD), and panoptic models like YOLOP are commonly used for object detection, segmentation, and road understanding. LiDAR-based models like VoxelNet~\cite{zhou2018voxelnet} and PointRCNN~\cite{shi2019pointrcnn} process 3D spatial data. Localization and mapping leverage SLAM~\cite{wang2007simultaneous} techniques augmented with visual odometry and deep feature extractors. For trajectory forecasting and environmental modeling, models such as LSTMs, RNNs, and graph neural networks (e.g., VectorNet~\cite{gao2020vectornet}) are employed. Decision-making and motion control are informed by deep reinforcement learning (DRL), inverse reinforcement learning, and policy gradient methods. Finally, end-to-end autonomous driving stacks may use CNN-based architectures, encoder-decoder frameworks, or multimodal large models integrating vision, LiDAR, and navigation data.}

\subsection{Simulation Environments for Autonomous Vehicle Testing}
\new{
Simulation environments are critical for developing, testing, and validating AV systems across diverse domains. According to the LLM's output, land-based simulation platforms such as CARLA~\cite{Dosovitskiy17}, AirSim (ground mode)~\cite{shah2017airsim}, Webots~\cite{Webots}, and CoppeliaSim~\cite{coppeliaSim} enable realistic testing of self-driving cars and wheeled robots with customizable sensors and traffic agents. Airborne systems benefit from tools like AirSim (drone mode), Gazebo (with PX4/RotorS)~\cite{koenig2004design}, and FlightGear~\cite{perry2004flightgear}, supporting UAV simulation and flight dynamics. Surface-sea and underwater vehicles are modeled using Gazebo plugins (usv\_sim\_lsa and uuv\_simulator), MATLAB's AUV toolbox, and commercial marine platforms. For space applications, NASA's Trick~\cite{penn2016trick} and Gazebo are the most  widely used simulators, as reported by the LLM.}

\section{Methodology}
\label{sec:methodology}

\begin{figure*}[t]
    \centering
    \includegraphics[width=1\textwidth]{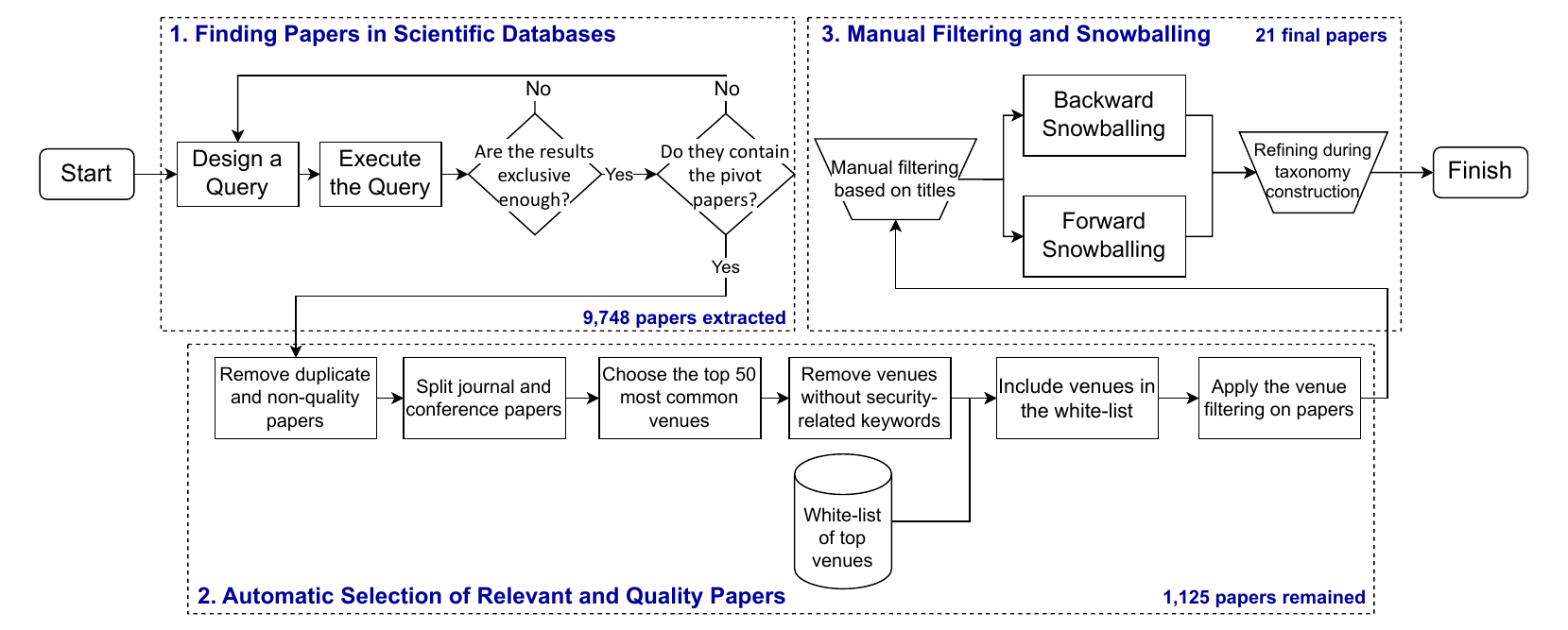}
    \caption{Overview of the paper collection process} \label{fig:overview}
\end{figure*}

Figure~\ref{fig:overview} presents an overview of our paper collection process. We adopted a systematic bottom-up approach, to curate a comprehensive collection of papers from diverse sources. We designed a search query to gather the papers from scientific databases such as Scopus (see below), spanning a 7-year period from 2017 to 2024\footnote{We selected 2017 as our starting year since research in the security of deep learning in AVs began to gain significant momentum after that year.}. This was followed by automatic selection and manual filtering. Automatic selection refines the results based on venue reputation, so as to include only well-established sources, while manual filtering involves the examination of each paper by multiple assessors. Additionally, we included relevant papers discovered through snowballing, ensuring a detailed collection by leveraging references to/from other works.

\subsection{Initial Paper Collection Through Scientific Databases}
\label{sec:initial_paper_colletion}

We employed \texttt{Findpapers}~\cite{grosman2020findpapers} to collect our initial set of papers. This tool allows us to design a customized search query, focusing on paper titles and abstracts, and supporting multiple databases including IEEE Explore\footnote{\url{https://ieeexplore.ieee.org/Xplore/home.jsp}}, Scopus\footnote{\url{https://www.elsevier.com/products/scopus}}, and ACM Digital Library\footnote{\url{https://dl.acm.org/}}. To ensure both comprehensive coverage and relevance, we iteratively refined our query based on two key criteria: the number of papers collected and the inclusion of three manually selected pivotal papers. Our pivotal papers contain two survey papers~\cite{biggio2018wild, cina2023wild} 
and DeepBillboard~\cite{zhou2020deepbillboard}, which deal with adversarial attacks on AVs\footnote{These papers are excluded from the final taxonomy since they do not satisfy criteria introduced in later phases of the process (e.g., no survey; system-level failure propagation).}. Through this iterative process, we balanced the query output, encompassing key papers while maintaining a broad search scope.

Given our focus on system-level attacks on AVs, we designed a search query that requires the papers' metadata (abstract, title, and keywords) to contain three key terms: `autonomous', `vehicle', and `attack'. Considering that not all papers may use these terms explicitly, we expanded our query to include synonyms. For `autonomous', we include variations such as `automated', `driverless', `self-driving', and `unmanned', connected with the \texttt{OR} operator to account for any of these terms appearing in the metadata. Similarly, for `vehicle*', we incorporate terms like `car*', `drone*', `truck*', `marine', etc, also connected with \texttt{OR}. Note that the use of asterisks ensures coverage of plural forms.

To ensure that each metadata contains at least one key term, all the term groups and their connected synonyms are linked with the \texttt{AND} operator. \new{Consequently, our final query is as follows:}

\lstdefinelanguage{querylang}{
  morekeywords={AND,OR,NOT},
  keywordstyle=\bfseries,
  sensitive=false
}

\begin{lstlisting}[language=querylang,breaklines=true,basicstyle=\ttfamily\footnotesize,label={lst:query}]
(([autonomous] OR [self-driving] OR [driverless] OR [unmanned] OR [automated])
AND
([vehicle*] OR [car*] OR [drone*] OR [truck*] OR [tractor*] OR [submarine*] OR [boat*] OR [maritime] OR [robotaxi*] OR [marine] OR [space] OR [spacecraft] OR [agriculture] OR [agricultural] OR [delivery] OR [construction])
AND
([attack*] OR [adversarial]))
\end{lstlisting}

Executing this query across three databases from January 2017 to January 2024, we retrieved a total of \new{9,748 papers}. Note that terms like `machine learning', `deep learning' and `system-level' are not explicitly mentioned in this query as they might exclude too many relevant papers. However, the subsequent manual filtering process ensured that only papers in the scope of our study (i.e., system level attacks against deep learning models) were taken into consideration. Moreover, the term `robots' is included to incorporate dynamic and mobile autonomous robots, such as delivery robots.

Next, our task was to refine this extensive pool of papers by filtering out those considered irrelevant or lacking in quality. Initially, we removed duplicate papers. Subsequently, we evaluated the publication venues based on established academic standards, such as the presence of a rigorous peer-review process and their focus on topics relevant to security. Papers from venues that did not meet these criteria were excluded. Employing a \textit{venue frequency filter}, we reduced the number of papers from \new{9,748} to 1,125. This filter was designed by ordering journals and conferences separately by frequency of occurrence in the pool and then selecting the top 50 most prevalent venues. Out of them, we eliminated venues lacking security-related keywords in their titles. For example, communication-related journals and conferences, and multimedia journals and conferences are excluded. Recognizing that this approach may exclude reputable software engineering/security-focused venues, we created a white list comprising prominent venues, ensuring their inclusion after the filtering process. This venue white-list is available online at \url{https://zenodo.org/records/16407709}.

We continued by manually filtering the 1,125 papers, focusing solely on their titles, which narrowed them down to 83 papers. From there, we delved deeper, examining the abstracts and further refining our selection to 30 papers.

\subsection{Snowballing}
\label{sec:snowballing}
The snowballing~\cite{wohlin2014guidelines} phase is aimed at expanding the pool of relevant papers by following their connections with other relevant papers. We employed both backward and forward snowballing techniques on the $30$ papers obtained from the previous phase.

Backward snowballing involves examining the references cited within each of the selected 30 papers and including additional papers based on their titles and abstracts. Conversely, forward snowballing involves identifying papers that cite the selected 30 papers. To achieve this, we utilized Google Scholar\footnote{\url{https://scholar.google.com}} to locate the papers referencing each selected paper and considered as candidates to be relevant papers only those within the first two pages of the search results. We limited our selection to the first two pages because Google Scholar ranks results based on relevance. To prevent excessive iterations, we did not perform snowballing on the newly added papers. 
In the end, our collection expanded to 50 papers. However, further filtering was still required, since selecting papers based only on their titles and abstracts is not necessarily accurate. This further selection was performed during the process of paper reading and categorization for taxonomy construction.

\subsection{Creation of Taxonomy Categories}
\label{sec:taxonomy_construction}

In this phase, we have read the 50 papers collected after snowballing, with the goal of creating the taxonomy categories and excluding papers that do not meet the inclusion criteria. In fact, sometimes it is necessary to read a paper fully to realize that it cannot be included in the taxonomy.

The process of category definition for the taxonomization of the papers was bootstrapped with the categories already available from the literature of attacks on DL models and AVs. Initially, we adopted ten categories suggested by Biggio et al.~\cite{biggio2018wild} and Cinà et al.~\cite{cina2023wild}. Subsequently, during taxonomy construction, we widened the scope to include concepts explicitly associated with system-level attacks and failures. This involved factors such as the system targeted by the attack and the specific system failures encountered during the attacks. In total, we identified 12 top-level categories, each listed in Table~\ref{tab:categories} with the corresponding explanations. A detailed description of these categories, as well as the sub-categories, is carried out below, in Section~\ref{subsec:taxonomytree}.

\begin{table}
    
\centering

\caption{\new{Examples of Possible Values of Each Taxonomy Category of System-level Attacks}}
\label{tab:categories}

\begin{tabular}{m{.3\textwidth}|m{.4\textwidth}m{.2\textwidth}}

\toprule

\rowcolor[HTML]{FFFFFF} \cellcolor[HTML]{FFFFFF}\textbf{Category }& \cellcolor[HTML]{FFFFFF}\textbf{Explanation}                           & \cellcolor[HTML]{FFFFFF}\textbf{\new{Possible} Values} \\

\midrule

\rowcolor[HTML]{FFFFFF}

\textbf{Application Domain}                                     & 
The domain or type of vehicle in which the attacker is attempting to cause a system-level failure
& Cars, Drones, \new{Trucks, Ships} \\ \cline{1-1}
\rowcolor[HTML]{EFEFEF} 
 \textbf{DL Model Under Attack}                                  & 
The specific deep-learning model \new{and module,} targeted by the attack
& Yolov5, ResNet-34, Dave2                                      \\ \cline{1-1}
\rowcolor[HTML]{FFFFFF} 
\textbf{System Under Attack}                                    & 
The specific autonomous system targeted by the attack                                      & Baidu Apollo in LGSVL, CARLA's agents \new{(MSF or SSF)}                \\ \cline{1-1}
\rowcolor[HTML]{EFEFEF} 
 \textbf{Attack Scenario}                                  & 
The particular scenario that must be present in order for the attack to be successful                & Cars in traffic jams, Cars in intersections                  \\ \cline{1-1}
\rowcolor[HTML]{FFFFFF} 
 \textbf{Attacked Target}                                        & 
The specific object or element that the attacker targeted to cause a failure                & Roads, Billboards, Input image                              \\ \cline{1-1}
\rowcolor[HTML]{EFEFEF} 
 \textbf{Attacker's Capability}                                  & 
The capabilities  the attacker needs to execute the attack                              & Attach stickers on the road, Install malware               \\ \cline{1-1}
\rowcolor[HTML]{FFFFFF} 
 \textbf{Attack Type}                                        & 
Whether the attack involves poisoning vs evasion tactics                                    & Poisoning, Evasion                                                                                      \\ \cline{1-1}
\rowcolor[HTML]{EFEFEF} 
\textbf{Attacker's Knowledge}                                   & 
The level of knowledge or access the attacker must possess about the system or model          & White-Box, Black-box, Gray-box                              \\ \cline{1-1}
\rowcolor[HTML]{FFFFFF} 
\textbf{Attack/Error Specificity}                               & 
Whether  attack/model-level error are generic or specific             & Generic/Specific Attack/Error \\ \cline{1-1}
\rowcolor[HTML]{EFEFEF} 
\textbf{Failure Specificity}                           & 
Whether the attacker aims to induce a specific system-level failure or not                & Generic, Specific                                                                                       \\ \cline{1-1}
\rowcolor[HTML]{FFFFFF} 
\textbf{Model-level Result}                                    & 
Outcome of the attack at the DL model-level                                        & Misclassification, Misdetection                            \\ \cline{1-1}
\rowcolor[HTML]{EFEFEF} 
\textbf{System-level Result}                                   & 
Impact of the attack at the system level                                   & Collision with object, Crash, Freeze                    \\ 

\bottomrule

\end{tabular}

\end{table}

In the phase of paper categorization and taxonomy refinement, we read each paper thoroughly. Specifically, each paper was read and categorized by at least two assessors. Weekly consensus meetings were held to resolve and negotiate any differences in the categorization. We excluded papers that do not specifically target the DL component of the AV under attack. 
For instance, papers that attack the hardware sensors and cause them to malfunction, resulting in incorrect point cloud generation and subsequent failure to recognize targets, are discarded as they do not directly involve any DL model~\cite{cao2023you}. Additionally, papers that only evaluate their attack on a dataset without executing the system, or collect their own dataset using cameras mounted on non-autonomous, manually driven vehicles, are excluded~\cite{zhou2020deepbillboard}.\new{ This criterion led to the exclusion of many papers, including the only two studies that specifically targeted Maritime Autonomous Surface Ships (MASS)~\cite{lee2023eval, lee2023vuln}.}
Papers that do not report any system-level failures, despite executing their attacks against AVs~\cite{fu2022ad,muller2022physical, huang2024spot}, are also discarded. Although these papers may report mispredictions of the vehicle's DL model, they do not provide information on the system-level vehicle's behavior once the attack has caused a model-level misprediction. As a result of this strict refinement process, the number of relevant papers was reduced from $50$ to \new{$21$}. \new{Other avenues for understanding attacks on AVs include research that targets network-based functionalities or attacks on DL models without system-level evaluation, which were not our goal in this taxonomy.}

\section{Results}
\label{sec:results}
In this section, we first provide a summary of each paper. Following this, we present the taxonomy mapping --- a table that classifies each paper according to the identified categories (multiple categories can be applied to each paper at the same time), offering a comprehensive overview of the existing research in the field. Finally, we comment on the taxonomy tree, which visually links each category and subcategory to the corresponding papers.

\subsection{Summary of papers}
A summary of the \new{21} papers that met our criteria is provided here to give an overview before delving into the detailed classification of each paper. The papers are organized in chronological order, starting with the most recent to the oldest.

\noindent \textbf{``Slowtrack: Increasing the latency of camera-based perception in autonomous driving using adversarial examples''}:
Ma et al.~\cite{ma2024slowtrack} introduced SlowTrack, a framework designed to increase the execution time of camera-based autonomous vehicle perception by inserting fake bounding boxes. This attack targets both object detection and tracking systems, resulting in vehicle crashes with a success rate of 95\%.

\new{
\noindent \textbf{``RPAU: Fooling the Eyes of UAVs via Physical Adversarial Patches''}:
Liu et al.~\cite{liu2024rpau} placed adversarial patches in the environment, including on moving vehicles, to disrupt drone perception. These patches caused the drones to either miss real objects and crash into them, change direction and veer off course, or detect fake objects and freeze in response.

\noindent \textbf{``Adversarial Attacks on Adaptive Cruise Control Systems''}:
Guo et al.~\cite{guo2023adv} used leading vehicles with adversarial patches placed on their rear, causing the target vehicle to either fail to detect the vehicle ahead or perceive it as being farther away -- potentially leading to a collision. To make the attack stealthy, they disguised the patch as a fake advertisement displayed on the back of a truck.
}

\noindent \textbf{``Learning when to use adaptive adversarial image perturbations against autonomous vehicles''}:
Yoon et al.~\cite{yoon2023learning} worked on injecting malware into the system to manipulate the input image. By adding perturbations, the attacker can create fake bounding boxes in specific locations, deceiving the victim's vehicle. This causes the vehicle to lose track of its intended target, potentially leading to a collision.

\noindent \textbf{``Deepmaneuver: Adversarial test generation for trajectory manipulation of autonomous vehicles''}:
Stein et al.~\cite{von2023deepmaneuver} successfully manipulated the steering angle prediction model of a victim's vehicle by installing a perturbation on a billboard on the left-hand side of the road. This caused the model to predict the wrong steering angle for a given input image, leading the vehicle to leave the road or collide.

\noindent \textbf{``Kidnapping deep learning-based multirotors using optimized flying adversarial patches''}:
Hanfeld et al.~\cite{hanfeld2023kidnapping} proposed an attack method capable of hijacking multi-rotor drones using flying adversarial patches, or small images strategically positioned in the surroundings. As a result, the drone deviates from its intended trajectory, losing track of the target human and instead following the adversarial patch.

\noindent \textbf{``Does physical adversarial example really matter to autonomous driving? Towards system-level effect of adversarial object evasion attack''}:
Wang et al.~\cite{wang2023does} explored the limitations of existing object evasion attacks in autonomous driving systems, noting their failure to achieve system-level effects due to inadequate understanding of the vehicle's physical properties. These include parameters like maximum/minimum acceleration/deceleration, steering rates, and sensor placements. To overcome these limitations, the paper introduced SysAdv, a novel attack design that integrates such system-specific knowledge. SysAdv significantly enhanced the success rate of attacks by approximately 70\%.

\noindent \textbf{``On data fabrication in collaborative vehicular perception: Attacks and countermeasures''}:
Zhang et al.~\cite{zhang2023data} proposed an attack method that exploits LiDAR-based collaborative perception in AVs, which can spoof or remove objects at specified locations in the victim’s perception results, making all mainstream types of collaborative perception schemes vulnerable.

\noindent \textbf{``Rolling colors: Adversarial laser exploits against traffic light recognition''}:
Yan et al.~\cite{yan2022rolling} proposed an attack targeting the color recognition systems of traffic lights in autonomous vehicles (AVs). They exploited the vulnerability of rolling shutters in CMOS cameras by directing a laser beam to create a colored stripe in the input image. This manipulation resulted in the misclassification of the traffic light's color.

\noindent \textbf{``Stop-and-go: Exploring backdoor attacks on deep reinforcement learning-based traffic congestion control systems''}:
Wang et al.~\cite{wang2021stop} researched poisoning a DRL-based controller in congested traffic. The attacker's car sets a trigger based on a combination of speed and position, causing the victim's vehicle to perform inappropriate actions like accelerating or decelerating incorrectly. This ultimately results in a collision with the vehicle in front, occurring on both one-lane and two-lane roads.

\noindent \textbf{``Attack and fault injection in self-driving agents on the carla simulator–experience report''}:
Piazzesi et al.~\cite{piazzesi2021attack} modified the trained agent's neurons and weights, as well as the input image, to cause the car to make incorrect steering decisions and mis-detect traffic lights. This resulted in collisions with surroundings, lane departure, going off-road, running through crossroads, and ignoring traffic lights.

\noindent \textbf{``Dirty road can attack: Security of deep learning based automated lane centering under Physical-World attack''}:
To deceive the victim vehicle's automated lane-centering system, Sato et al.~\cite{sato2021dirty} printed specially designed patches on the road that resembled dirty road markings. This manipulation caused the system to make incorrect steering predictions, leading the vehicle to engage in hazardous maneuvers such as turning left or right incorrectly, driving off-road, and causing collisions.

\noindent \textbf{``Invisible for both camera and lidar: Security of multi-sensor fusion based perception in autonomous driving under physical-world attacks''}:
Cao et al.~\cite{cao2021invisible} conducted a physical sensor-based attack aimed at deceiving the camera and Lidar perception systems of the victim's vehicle, specifically while driving in a single road lane. Their strategy involved 3D printing and placing an adversarial traffic cone on the road that was designed to appear broken and thus remain undetected by both sensors. This deception caused the vehicle's systems to fail to recognize the object, resulting in collisions.

\noindent \textbf{``Too good to be safe: Tricking lane detection in autonomous driving with crafted perturbations''}:
Jing et al.~\cite{jing2021too} implemented a technique where three white dots are painted on the road either upon completing the road lines or when there is space between them. These dots are designed to trick cars into mistaking them for real lane markers, leading the vehicles to follow the dots and veer into the opposite lane.

\noindent \textbf{``Robust roadside physical
adversarial attack against deep learning in lidar perception modules''}:
Yang et al.~\cite{yang2021robust} utilized 3D printing to create a small adversarial object, which they placed on the road. The purpose was to deceive the victim's car into mistaking the object for another vehicle, prompting it to either change lanes or come to a complete stop upon detecting the object.

\noindent \textbf{``ML-driven malware that targets AV safety''}:
Jha et al.~\cite{jha2020ml} installed malware to introduce noise to the outputs of object detection, tracking, and sensor fusion systems. The malware modifies the outputs under specific conditions by adding perturbative data as sensor noise. This noise makes the victim's car believe that a car in another lane is moving into its lane or that a car in the same lane in front of it has disappeared. Additionally, it can make the car think that pedestrians are moving into the street. This results in emergency braking or unwanted acceleration, leading to a car crash.

\noindent \textbf{``Attacking vision-based perception in end-to-end autonomous driving models''}:
In the research of Boloor et al.~\cite{boloor2020attacking}, when the victim's car is at an intersection or on a turning road, the attacker paints adversarial road lines pointing in the opposite direction of the lane turn, causing the car to turn the wrong way.

\noindent \textbf{``Feasibility and suppression of adversarial patch attacks on end-to-end vehicle control''}:
Pavlitskaya et al.~\cite{pavlitskaya2020feasibility} proposed manipulating an autonomous vehicle's trajectory by placing a printed adversarial patch on the roadside. This action causes the vehicle to steer towards the patch, resulting in a collision.

\noindent \textbf{``Phantom of the ADAS: Securing advanced driver-assistance systems from split-second phantom attacks''}:
Nassi et al.~\cite{nassi2020phantom} developed a method to project images of humans or traffic signs onto roads or flat surfaces. These projections, referred to as phantoms, are intended to deceive cars into perceiving them as real signs or pedestrians, prompting the vehicles to respond accordingly.

\noindent \textbf{``Adversarial sensor attack on lidar-based perception in autonomous driving''}:
Cao et al.~\cite{cao2019adversarial} directed lasers from an adversarial vehicle toward the target vehicle's Lidar system. This intentional interference aims to create false objects that do not exist, leading the car to freeze or suddenly brake.

\noindent \textbf{``Trojaning attack on neural networks''}:
In the work of Liu et al.~\cite{liu2017trojaning}, the attacker deceived the victim's vehicle by poisoning its driving model with trojan data to cause a wheel angle misprediction. When the vehicle encounters the trojan attached to a billboard, it is instructed to turn right, leading to the vehicle driving off the road.

\subsection{Mapping the papers}

\label{subsec:mapping}
For the taxonomy mapping, we employed a rigorous process: each paper was independently reviewed and categorized by at least two assessors. To ensure consistency and reliability, the assessors regularly met to discuss and resolve disagreements in their mappings. The results of this process are presented in the Appendix, in Tables~\ref{tab:taxonomy1}, \ref{tab:taxonomy2}, and \ref{tab:taxonomy3}, where each column represents a distinct category, and each row corresponds to an individual paper and its associated categorization. 

We conducted weekly meetings to ensure that the assessors' perspectives aligned with each other while reading and categorizing the papers. Over time, the meetings led to a decrease in disagreements, which typically involved the category to choose (major disagreement) or how to precisely formulate a (sub-)category (minor disagreement). 
For paper mapping, we used a shared spreadsheet containing 228 cells.
Among the 228 cells that were filled, around 20 cells (9\%) faced major disagreements and approximately 80 cells (35\%) faced minor disagreements. On average, 8 disagreements were resolved per meeting, resulting in a total of about 96 resolved disagreements over 12 meetings. Initially, the categories with the most disagreement were ``Attacker's System/Model Knowledge,'' followed by ``Attack/Error Specificity'' and ``Attacked Target.'' Possible reasons for the higher disagreement in these categories were the 
difficulty of giving a unique interpretation and purpose to these categories, and the difficulty of extracting the needed information from some papers, as it is not always directly provided by the authors. 
The consensus meetings turned out to be a fundamental instrument to address such initial disagreements and to come out with a common understanding and interpretation, which was reflected in a much clearer definition of the categories and choice of the labels for the categories after such meetings.
\subsection{Taxonomy tree} \label{subsec:taxonomytree}

\begin{sidewaysfigure}
    \centering
    \includegraphics[width=\textwidth]{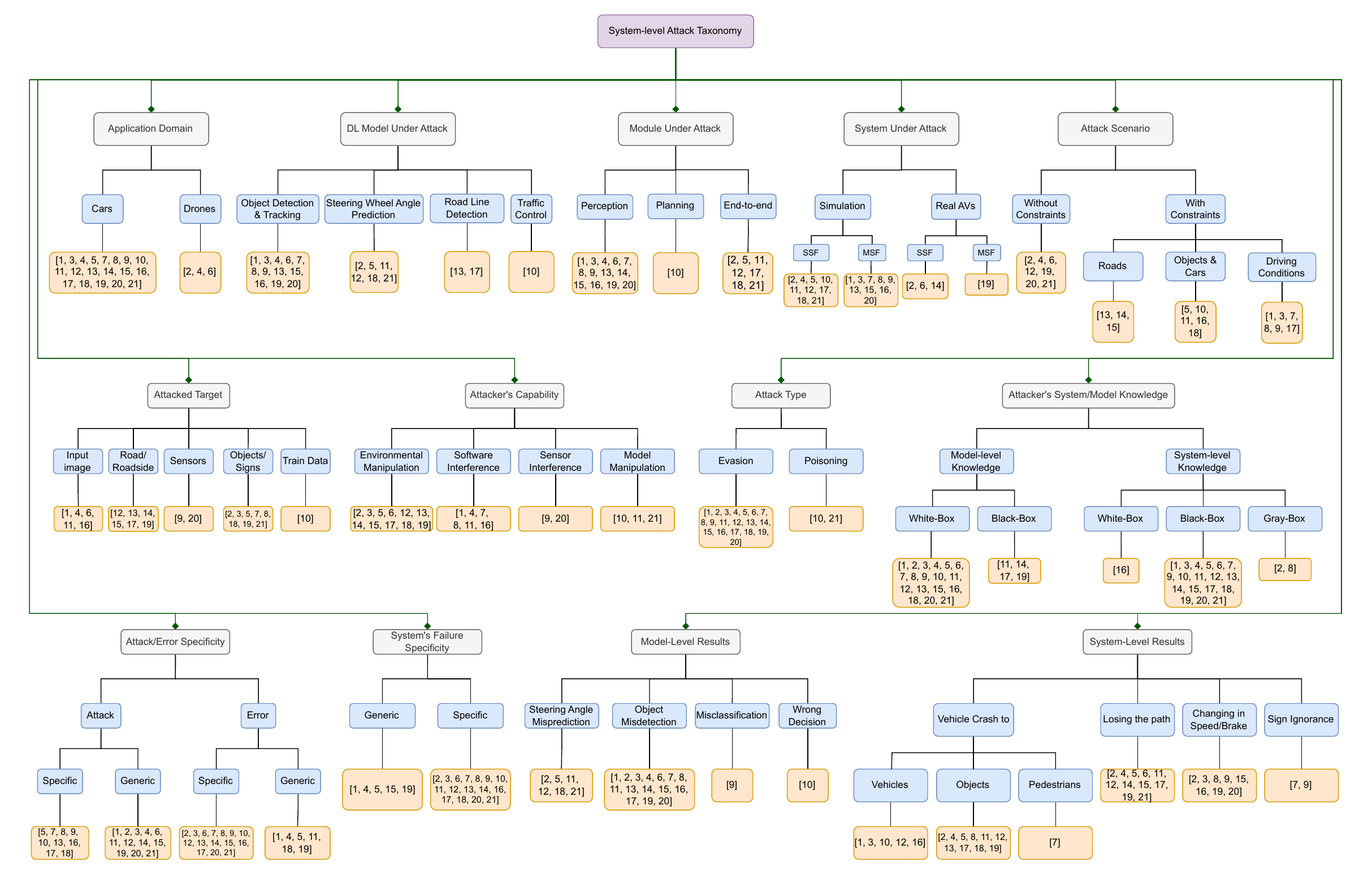}
    \caption{Taxonomy Tree} \label{fig:tree}
\end{sidewaysfigure}

Figure~\ref{fig:tree} provides a visual representation of the resulting taxonomy as a tree. 
The root node branches into 12 top-level categories, each further subdivided into relevant subcategories. The leaf nodes of the tree contain lists of papers associated with their respective (sub)categories. 
Each category and subcategory is detailed in the following. %

\begin{enumerate}
    \item \textbf{Application Domain:} Domain or vehicle type that is the target of the attack, such as ``Cars'' and ``Drones. \new{If we compare the application domains covered by the papers included in the taxonomy with the AV types reported by the LLM in Table~\ref{tab:AVoverview}, we can notice that most AV surfaces, including sea, underwater and space, are uncovered. On the other hand, all these uncovered AV types represent critical domains where security testing would be quite important. We conclude that research is missing on the system level security of AVs operating on surfaces other than land and air.}

    \item \textbf{DL Model Under Attack:} The DL model under attack, which is responsible for the system-level failure, such as YOLOv5, ResNet-34, or DAVE-2. The ``DL Model Under Attack'' category has four subcategories. The first, ``Object Detection \& Tracking'', includes DL models (e.g., YOLO) that handle object tracking or detection. The ``Steering Wheel Angle Prediction'' covers models that control the car's steering angle. ``Road Line Detection'' helps the system detect and follow road lines, while ``Traffic Control'' assists in navigating complex traffic situations. \new{All of these DL models contribute to the autonomous capabilities of vehicles and are part of a modular AV architecture, such as perception, planning, and end-to-end driving.}
    \new{If we compare the models under attack covered by the papers included in the taxonomy with the AV models and modules reported by the LLM in Table~\ref{tab:AVoverview}, we can notice that most types of modules are covered. However, modules for localization and mapping have not been subjected to system-level security testing in any of the included papers. This might be due to the difficulty associated with faithful simulation of such modules. Another missing module type is associated with the low level ``control'' functionality, often implemented via reinforcement learning. This indicates a general preference of security testing papers toward supervised learning, as opposed to reinforcement learning. A similar preference seems to hold for model-level adversarial attacks, which are typically crafted against neural networks trained in a supervised setting.}

    \item \textbf{System Under Attack:} The autonomous system under attack, such as a driving agent like Baidu Apollo in the LGSVL simulator or the driving agent in the CARLA simulator. ``System Under Attack'' has two subcategories, ``Simulation'' and ``Real AVs'', respectively indicating that the system under attack is simulated or is a real-world autonomous vehicle. ``Baidu Apollo''~\cite{apollo} and ``CARLA's Agent''~\cite{Dosovitskiy17} represent examples of autonomous systems implemented in the LGSVL~\cite{rong2020lgsvl} and CARLA simulators --- the most popular simulation environments in our list of papers. Other less popular simulators include BeamNG~\cite{beamng_tech} and Udacity~\cite{udacity}. ``Tesla'' is instead an example referred to by papers that used a real Tesla vehicle to test their attacks. \new{As systems in AVs may use either single or multi-sensor fusion to make autonomous decisions, in this category, we distinguish between them by using \textit{MSF} for multi-sensor fusion systems and \textit{SSF} for single-sensor fusion systems.}
    \new{If we compare the systems under attack covered by the papers included in the taxonomy with the simulation environments reported by the LLM in Table~\ref{tab:AVoverview}, we can notice that simulators for several AV surfaces, including sea, underwater and space, are not used in the included papers, or are used for another surface (e.g., Gazebo can be used to simulate air, sea, underwater and space, but in the included papers it was used only to simulate AVs operating in the air -- specifically, drones). Similarly to the remark made above for the application domain, the missing simulation environments point to the need for further research to bring system level attacks to simulators operating in the sea, underwater or in the space.}

    \item \textbf{Attack Scenario:} The scenario in which the attack is designed to be effective, such as when a car is at an intersection or stuck in a traffic jam. Hence, the ``Attack Scenario'' category provides details on the \textit{necessary conditions for an attack}, i.e., the conditions that must be satisfied in order for an attack to be successful. Some papers did not specify any constraints, so they fall under the ``Without Constraints'' subcategory. For those requiring specific conditions, we created the ``With Constraints'' subcategory, which includes three further distinctions. The ``Roads'' subcategory covers attacks that need modifications to the road itself. The ``Objects \& Cars'' subcategory includes attacks that require objects, such as billboards or non-autonomous cars, to be positioned in specific places. Lastly, the ``Driving Condition'' subcategory relates to papers that do not require physical changes but need the AV to be in a specific scenario, such as approaching a traffic light, being at an intersection, or navigating a turn.

    \item \textbf{Attacked Target:} The specific target of the attack, such as a billboard where to place adversarial noise or a road where to add adversarial lines. If an attack operates as malicious software that directly alters the input (e.g., image) received by the DL model, then the attacked target would be e.g. the input image. The subcategories of ``Attacked Target''  concern the specific elements that the attacker needs to change, manipulate, or perturb. This can include the ``Input Image'' fed into the DL model, the ``Road'' or ``Roadside'', the AV's ``Sensors'', ``Objects'' like billboards or traffic ``Signs'' like stop signs, and the ``Training Data'' used to train the model before deployment.

    \item \textbf{Attacker's Capability:} The capabilities required by the attacker to execute the attack, such as the ability to alter the physical environment or to install malware into the system. The ``Attacker's Capability'' category has four subcategories. Attacks requiring changes to the environment, such as placing objects on the road, fall under the ``Environmental Manipulation'' subcategory. Those needing alterations at the software level, like acting as a man in the middle, are categorized under ``Software Interference''. Attacks that interfere with the sensors' regular functionality, such as directing lasers at them, are included in the ``Sensor Interference'' subcategory. Lastly, those requiring manipulation of the model, such as adding backdoors to the training data, fall under the ``Model Manipulation'' subcategory.

    \item \textbf{Attack Type:} The attacker's general approach, such as altering training data (``Poisoning'') or producing mis-prediction at inference time (``Evasion'').
    
    \item \textbf{Attacker's Knowledge:} The attacker's knowledge about the system or model, split into white-, black-, or grey-box knowledge:
    \begin{itemize}
        \item ``Model-level White-Box'': Full access to the code used to create the model and to the dataset used to train/test the model.
        \item ``Model-level Black-Box'': The attacker can only know the input and output vectors accepted/produced by the model.
        \item ``System-level White-Box'': The attacker has unrestricted access to the entire system, including the source code of the system, \new{the hardware specs}, DL models, simulator (if any) and datasets.
        \item ``System-level Black-Box'': The attacker's knowledge is limited to input (e.g., sensor) and output (e.g. actuator) data, with no visibility into the internal system components.
        \item ``System-level Grey-Box'': The attacker has some insights into the system but not to the full extent of a white-box setting (e.g., the code of some components is available, but not the code of the entire system).
    \end{itemize}

    \item \textbf{Attack/Error Specificity:} \new{This category includes Attack-Specific or Attack-Generic, as well as Error-Specific or Error-Generic}. Some attacks are highly specific, targeting only one particular object or aspect (e.g., only traffic signs are attacked), which makes them ``Attack-Specific''. On the other hand, the ``Attack-Generic'' subcategory represents a more general type of attack, which can be applied to a wider range of targets (e.g., any object on the side of the road). Regarding the model mis-predictions resulting from the attack, ``Error-Specific'' attacks aim to induce a particular output or trigger unique misbehavior in the model (e.g., confusing a 60 speed limit with a 90), while ``Error-Generic'' attacks allow any kind of misbehavior of the model (e.g., mis-classifying a traffic sign in any possible way).

    \item \textbf{System's Failure Specificity:} Similar to the previous category, this one indicates whether the attacker expects a particular system-level failure. For instance, if the attacker aims to make the vehicle suddenly brake, the attack is considered ``Failure-Specific''. Alternatively, a ``Failure-Generic'' attack aims at a general system failure, without specifying the exact expected outcome.
    
    \item \textbf{Model-level Results:} The outcome of the DL model resulting from the attack, such as a mis-classification or mis-detection by the object classifier or object detection component of the autonomous vehicle. In this category, papers causing a wrong steering angle prediction fall under the ``Steering Angle Misprediction'' subcategory. Those causing the object detection module to fail at detecting objects correctly are categorized under ``Object Misdetection''. The ``Misclassification'' subcategory includes e.g. a paper where traffic light colors are incorrectly classified. Finally, the ``Wrong Decision'' subcategory is for instances where models make incorrect decisions, such as in traffic jams.

    \item \textbf{System-level Results:} This category indicates the outcomes of the attack at the system level, such as collisions with other cars, collisions with the environment, or sudden braking, resulting from model-level misbehavior propagated into system-level faults. For the ``System-Level Results'', we have four subcategories. Some papers cause AV cars to crash into various things. Therefore, under the ``Vehicle Crash To'' subcategory, we have three more subcategories to specify what the AV crashed into. If the AV crashes into another vehicle (often a non-autonomous car), it falls under the ``Vehicles'' subcategory. The ``Objects'' subcategory is for instances when the AV hits objects such as road curbs, billboards, traffic cones, etc. There's also a subcategory for cases where the AV crashes into ``Pedestrians''. For results that don't fall under ``Vehicle Crash To'' we have additional subcategories. The ``Losing the Path'' subcategory covers cases where the AV loses its path and goes into another lane or off-road. The ``Changing in Speed/Brake'' subcategory includes attacks that cause the AV to accelerate, decelerate, or suddenly brake. Lastly, the ``Sign Ignorance'' subcategory is for instances where the AV ignores stop signs or traffic lights.
\end{enumerate}

\section{Discussion}
\label{sec:discussion}
In this section, we first answer the research questions and then point out to the takeaways of this work and the future research directions.

\begin{figure}
\centering
\begin{subfigure}{0.33\linewidth}
    \centering
    \includegraphics[width=1\linewidth]{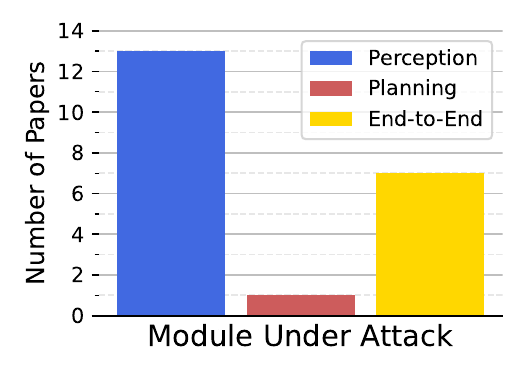}  
    \caption{}
    \label{subfigure:ModuleUn}
\end{subfigure}
\begin{subfigure}{0.33\linewidth}
    \centering
    \includegraphics[width=1\linewidth]{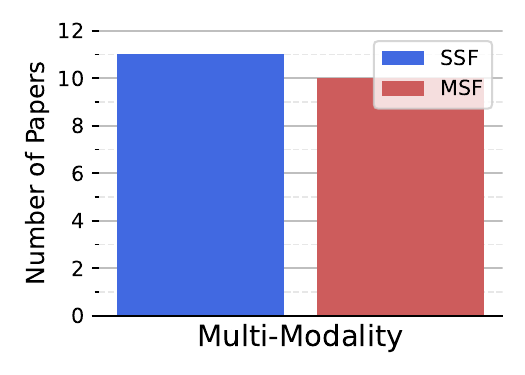}  
    \caption{}
    \label{subfigure:MultiMod}
\end{subfigure}
\begin{subfigure}{0.33\linewidth}
    \centering
    \includegraphics[width=1\linewidth]{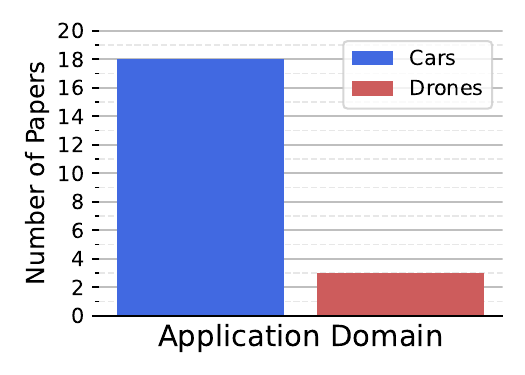}  
    \caption{}
    \label{subfigure:ADomain}
\end{subfigure}
\begin{subfigure}{0.33\linewidth}
    \centering
    \includegraphics[width=1\linewidth]{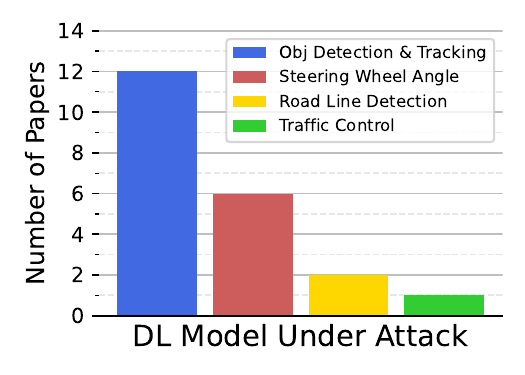}  
    \caption{}
    \label{subfigure:DLModel}
\end{subfigure}
\begin{subfigure}{0.33\linewidth}
    \centering
    \includegraphics[width=1\linewidth]{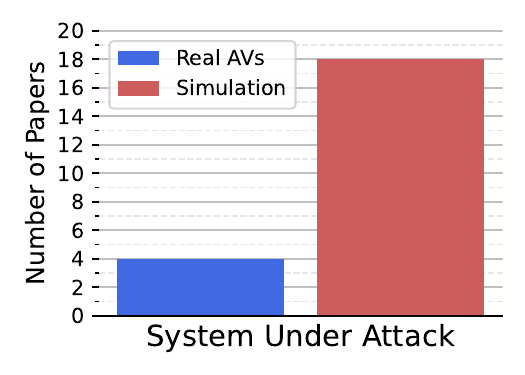}  
    \caption{}
    \label{subfigure:SystemUn}
\end{subfigure}
\begin{subfigure}{0.33\linewidth}
    \centering
    \includegraphics[width=1\linewidth]{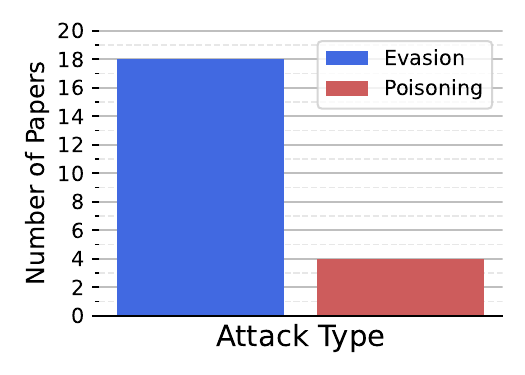}  
    \caption{}
    \label{subfigure:Type}
\end{subfigure}
\begin{subfigure}{0.33\linewidth}
    \centering
    \includegraphics[width=1\linewidth]{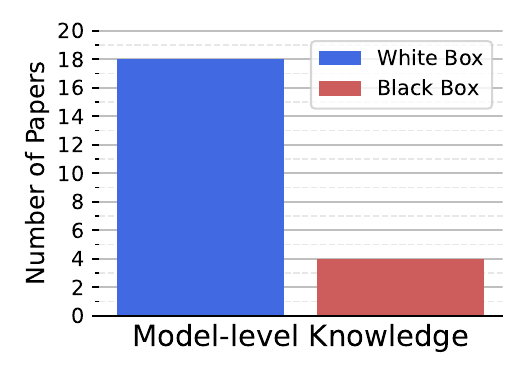}
    \caption{}
    \label{subfigure:ModKnowledge}
\end{subfigure}
\begin{subfigure}{0.33\linewidth}
    \centering
    \includegraphics[width=1\linewidth]{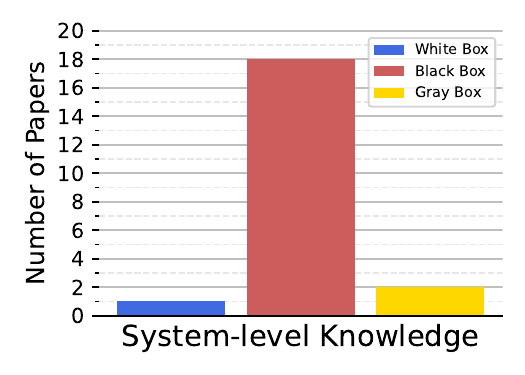}
    \caption{}
    \label{subfigure:SysKnowledge}
\end{subfigure}
\begin{subfigure}{0.33\linewidth}
    \centering
    \includegraphics[width=1\linewidth]{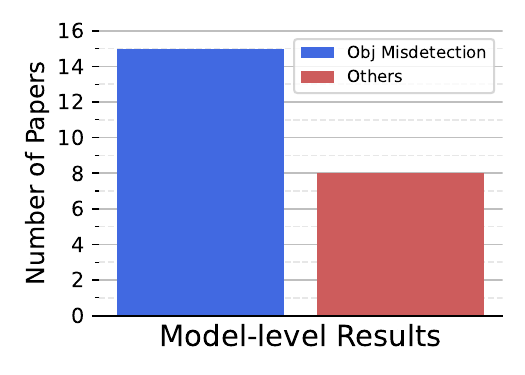}  
    \caption{}
    \label{subfigure:ModResults}
\end{subfigure}
\begin{subfigure}{0.33\linewidth}
    \centering
    \includegraphics[width=1\linewidth]{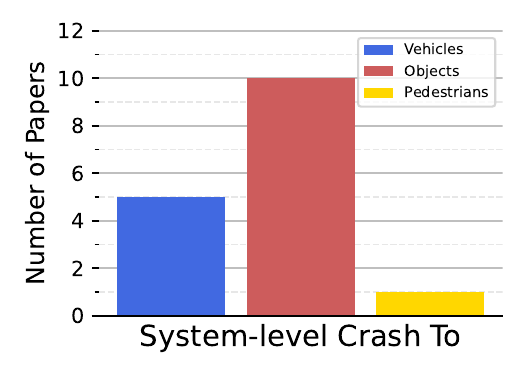}  
    \caption{}
    \label{subfigure:CrashTo}
\end{subfigure}
\caption{Paper distribution in the taxonomy tree}
\label{fig:distribution}
\end{figure}

\subsection{Answers to Research Questions}

\subsubsection{RQ1 [Attack features]}

Figure~\ref{fig:distribution} illustrates the distribution of papers across the taxonomy tree. It reveals substantial differences in the number of papers associated with some categories.

\new{
Figure~\ref{subfigure:ModuleUn} illustrates that the perception module is the most frequently targeted in attacks~\cite{ma2024slowtrack,cao2019adversarial,yoon2023learning,hanfeld2023kidnapping, wang2023does, zhang2023data, yan2022rolling, cao2021invisible, jing2021too, yang2021robust, jha2020ml, nassi2020phantom, guo2023adv}. End-to-end models follow as the second most common target~\cite{liu2024rpau,von2023deepmaneuver,piazzesi2021attack,sato2021dirty, boloor2020attacking, pavlitskaya2020feasibility,liu2017trojaning}. However, only a single attack was recorded on the prediction and planning module, while no attacks were reported on the DL components of the localization or control modules. Since the perception module is the first to process sensor data in an AV, any compromise at this stage can lead downstream modules to make incorrect decisions, ultimately resulting in faulty control outputs. End-to-end models are also highly vulnerable, as they bypass preprocessing and act directly on raw sensor inputs, increasing their susceptibility to adversarial manipulation.

Figure~\ref{subfigure:MultiMod} shows that attacks are fairly evenly distributed between MSF and SSF systems. Although MSF systems are often considered more secure due to their integration of multiple sensor types~\cite{lou2022testing}, this study reveals that they remain vulnerable to well-crafted adversarial attacks.
}

In Figure~\ref{subfigure:ADomain}, we observe that only three papers focus on the application domain of autonomous drones, compared to 18 papers on autonomous cars. 
\new{As autonomous drones continue to gain traction across various sectors, we expect such proportion to change in the future. The notable lack of research on system-level adversarial attacks in other application domains, such as autonomous trucks, buses, ships, spacecraft, and more, highlights a clear gap in the current literature, pointing to a whole research area that could be investigated more deeply in the future.}

Figure~\ref{subfigure:DLModel} shows that most attacks target the object detection and tracking models of AVs, indicating that this component is the most vulnerable part of an AV system. Steering wheel angle prediction was also a notable target for attacks, as it is influenced by numerous factors on the road that can be the target of an attack~\cite{von2023deepmaneuver, piazzesi2021attack, sato2021dirty, pavlitskaya2020feasibility, liu2017trojaning}. 

Figure~\ref{subfigure:SystemUn} clearly shows that the majority of papers focused on deploying attacks in a simulation environment. This trend may be due to the inherent dangers of the attacks, or legal constraints in various countries, making si\-mu\-lation environments more feasible for conducting such research. 

Figure~\ref{subfigure:Type} illustrates the approach chosen by attackers to deploy their attacks. Evasion attacks are prevalent, likely because poisoning attacks require access to the model's dataset before training. This scenario is hard to achieve in the real world, making evasion attacks more realistic and feasible.

In Figure~\ref{subfigure:ModKnowledge}, the model-level knowledge is predominantly considered to be white-box. While this may not be very practical in real-world scenarios, it is often necessary for attackers to make their attacks effective. In this respect, the four papers~\cite{piazzesi2021attack, jing2021too, boloor2020attacking, nassi2020phantom} that utilized a black-box knowledge approach would represent more feasible options for real-world attacks. 

Figure~\ref{subfigure:SysKnowledge} addresses the system-level knowledge required for attacks. It is noteworthy that almost every attack does not require in-depth knowledge or access to the source code of the system. 
On one hand, this might be due to the practical difficulties that prevent the acquisition of white-box or grey-box knowledge of complex, possibly proprietary, systems. On the other hand, while at the model level, white-box knowledge is often a prerequisite that makes the attack otherwise impossible, at the system level, plain knowledge of the environment elements and system input is often enough to introduce manipulations that cause a model mis-prediction and possibly a system failure.

In line with Figure~\ref{subfigure:DLModel}, Figure~\ref{subfigure:ModResults} also shows that the outcome of attacks on the DL component is predominantly the misdetection of targets or objects the AV is trying to recognize. These objects can include traffic signs, cars, pedestrians, etc. This highlights a significant vulnerability in the object detection and recognition capabilities of autonomous vehicles. 

Figure~\ref{subfigure:CrashTo} suggests that most crashes involving AVs were with objects such as obstacles on the road. The figure indicates that only one paper reported crashes involving pedestrians. This could be related to the difficulty in effectively targeting pedestrians as part of an attack.
\\
\begin{mdframed}[backgroundcolor=cyan!15]
\textbf{RQ1 [Attack Features]}: \new{The distribution of papers across the taxonomy categories is notably unbalanced. System-level attacks on AVs are primarily focused on cars and, to a lesser extent, drones. Other types of AVs, such as tractors, spacecraft, and maritime surface ships, are largely absent from the current research landscape. Most attacks target the \textit{perception} module of AVs, with significantly less attention given to other critical components like \textit{planning}, \textit{localization}, and \textit{control}. While one might assume that AVs employing MSF are more resilient to adversarial attacks, this taxonomy reveals that MSF systems are still vulnerable. The majority of studies focus on simulated car environments, targeting components such as object detection/tracking and steering angle prediction, with the primary goal being mispredictions (evasion). These attacks are often conducted with white-box access to the model but only black-box knowledge of the system. At the model level, attacks are typically designed to cause object misdetection, whereas at the system level, the aim is often to cause the AV to crash into other vehicles or obstacles.}

\end{mdframed}

\subsubsection{RQ2 [Attacked Components]}

By examining the taxonomy tree, we observe that the DL models under attack consist of object detection \& tracking, steering angle prediction, lane detection, and traffic control models. 
Most of them support the perception tasks carried out by the AV, by identifying relevant features in the perceived input and manipulating them to cause a mis-prediction. Often, the perceived input is a camera image~\cite{ma2024slowtrack, yoon2023learning, hanfeld2023kidnapping, piazzesi2021attack, jha2020ml}, making it possible to reuse the wide range of attacks investigated in previous works on model-level evasion against images, achieved via adversarial noise patches~\cite{biggio2018wild}. 
This is confirmed by the knowledge required by these attacks, which is white-box at the model level and black-box at the system level (see Figures~\ref{subfigure:ModKnowledge}, \ref{subfigure:SysKnowledge}).
Hence, the problem becomes how to manipulate the simulation environment in which the AV operates in order to introduce the attack patches that are known to trigger a mis-prediction at the model level, assuming that the induced error will propagate inside the system, which is known only as a black box, to eventually cause a system failure.

The systems under attack include Baidu Apollo, CARLA's driving agent, Tesla, and several other AV systems. However, most attacked systems are executed within a simulator (see Figure~\ref{subfigure:SystemUn}). There are several practical reasons to prefer simulated AVs over real ones when experimenting with new attacks against AVs, including: (1) simulations are cheaper, faster and not affected by safety constraints; (2) simulations are less constrained and offer higher flexibility in the manipulation of the environment, which could be very expensive or infeasible in the real world; (3) simulations provide more information about the execution, e.g., obtained by monitoring and logging different types of execution/simulation data, which is useful to fine-tune an attack. Such information might be unavailable in the real world. \\

\begin{mdframed}[backgroundcolor=cyan!15]
\textbf{RQ2 [Attacked Components]}: The most attacked components are image-processing neural networks \new{in the perception module}, for which several model-level attacks are readily available in the literature. The problem is then how to manipulate the environment such that the model-level input resembles an adversarial input and how to ensure the propagation of a model-level error to a system-level failure. This seems easier to achieve in simulation than in the real world because a simulator offers higher flexibility, more information, and is generally cheaper, faster, and not dangerous for human beings. \new{Finally, almost every attack on MSF systems required a white-box knowledge of the DL model.}
\end{mdframed}

\subsubsection{RQ3 [Threat Models]}

The most common assumption is that the attacker has white-box access to a DL model used in an AV system, while the system itself is known only as a black-box. Under the assumption of white-box model knowledge and black-box system knowledge, the papers included in our taxonomy explored various threat models: 

\begin{enumerate}
\item \textbf{Outside Attack}: The attacker cannot manipulate the DL model input. Hence, the attacker can only manipulate the environment~\cite{liu2024rpau, von2023deepmaneuver, hanfeld2023kidnapping, sato2021dirty, cao2021invisible, jing2021too, yang2021robust, boloor2020attacking, pavlitskaya2020feasibility, nassi2020phantom} or can interfere with the sensors~\cite{yan2022rolling,cao2019adversarial} to supply an adversarial input to the model.
\item \textbf{Inside Attack at Inference Time}: The attacker can manipulate directly the DL model input at inference time~\cite{ma2024slowtrack, yoon2023learning, wang2023does, zhang2023data,piazzesi2021attack,jha2020ml}. This can be achieved e.g. by installing some malware that can intercept and occasionally replace the actual model input with an adversarial one. 
\item \textbf{Inside Attack at Training Time}: The attacker can manipulate directly the dataset used to train the DL model~\cite{wang2021stop, piazzesi2021attack, liu2017trojaning}. This can be achieved e.g. by obtaining unauthorized access to the victim's computers or with the complicity of an employer working at the victim's company.
\end{enumerate}
\new{
Under the assumption of black-box model knowledge, nearly all attacks were \textbf{Outside Attacks}, manipulating the environment, such as the road, with adversarial input~\cite{jing2021too, boloor2020attacking, nassi2020phantom}. The only exception is an attack that employs a weight injection tool to generate perturbations directly on images~\cite{piazzesi2021attack}}

The above-listed threat models require an increasing amount of capabilities and permissions on the attacker's side, with the Outside Attack threats being the easiest to mount and the Inside Attack threats the most difficult and challenging. It should be noticed that not all papers are explicit about the underlying threat model they assume. As a result, we often had to infer it from the algorithmic description of the attack. As the risk associated with an attack depends not only on the damage it can cause, but also on the probability that an attacker can successfully mount the attack, explicit knowledge of the threat model is important to understand and possibly estimate the likelihood of an attack.
\\
\begin{mdframed}[backgroundcolor=cyan!15]
\textbf{RQ3 [Threat Models]}: Threat models span from limited capabilities of manipulation of the environment where an AV operates, to the possibility of directly corrupting or replacing input values at inference or even at training time. Such a wide range corresponds to different degrees of attack likelihood and realism. Papers often lack explicit information on the assumed threat model, which makes it difficult for the reader to assess the feasibility of an attack in a given context. \new{Although modifying the environment might appear easy, most attacks rely on white-box access to the DL model. This makes it less realistic to believe that such attacks could occur in real-world scenarios, unless the model is leaked or the attacker’s surrogate model closely resembles the target model. 
}
\end{mdframed}

\subsubsection{RQ4 [Consequences]}
    
Most attacks that cause a misprediction of the steering angle result in the vehicle losing its path and sometimes hitting road objects, such as side walls~\cite{von2023deepmaneuver, piazzesi2021attack, sato2021dirty,pavlitskaya2020feasibility, liu2017trojaning}. 
On the other hand, object misdetection often causes the vehicle to fail to see an object or to perceive a fake object as real, leading to unexpected changes in speed or crashes~\cite{ma2024slowtrack,guo2023adv,zhang2023data,piazzesi2021attack,cao2021invisible,yang2021robust,jha2020ml,boloor2020attacking,nassi2020phantom,cao2019adversarial}. 
Misclassifications by the model and incorrect decisions can cause the car to ignore traffic signs or collide with other vehicles~\cite{yan2022rolling,wang2021stop}.

There are two main chains of attack consequences from the input to a system failure:

\begin{enumerate}
\item \textbf{Targeted \new{Attacks}}: one or more adversarial inputs \new{are fed} into the DL model and cause a misprediction, which goes in a specific direction, aimed at producing a specific system-level failure~\cite{liu2024rpau,cao2019adversarial,zhang2023data,yan2022rolling,wang2021stop,piazzesi2021attack,sato2021dirty,cao2021invisible,jing2021too,jha2020ml,boloor2020attacking,pavlitskaya2020feasibility,nassi2020phantom,cao2019adversarial,liu2017trojaning}.  For instance, a patch is added to the environment in order to revert the numerical sign of the steering angle prediction, which is supposed to make the car go off its lane~\cite{boloor2020attacking}. 
\item \textbf{Untargeted \new{Attacks}}: one or more adversarial inputs \new{are fed} into the DL model and cause an arbitrary misprediction. Propagation to an arbitrary system failure is conjectured and often possible, but it is not targeted explicitly~\cite{ma2024slowtrack,yoon2023learning,von2023deepmaneuver,yang2021robust,nassi2020phantom}. For instance, perturbations are added to the input image to cause arbitrary, incorrect bounding boxes. As a result of the model-level misclassification, the self-driving car misbehaves~\cite{ma2024slowtrack}.
\end{enumerate}

Clearly, the targeted \new{attack} chain is more effective as it causes a model-level mis-prediction that is specifically functional to an overall misbehavior of the whole system. However, such targeted chains require deeper knowledge of the system, which is often unavailable or incomplete. Hence, often papers adopt the untargeted \new{attack} chain and then just observe and evaluate its effectiveness empirically, with no warranty of success.
\\
\begin{mdframed}[backgroundcolor=cyan!15]
\textbf{RQ4 [Consequences]}: There are two main approaches to achieve the system-level consequences of an attack: (1) specific model-level mis-predictions are targeted in specific execution contexts, to ensure the desired consequence at the system level; (2) arbitrary model-level mis-predictions are triggered, and system-level consequences are just observed, but not controlled.
\end{mdframed}

\subsection{Implications for Practitioners and Researchers}

Based on our analysis of the papers and based on the taxonomy extracted from them, we can distill a few lessons and implications that are potentially useful to practitioners and represent interesting opportunities for researchers.

\begin{enumerate}
    \item \textbf{Vulnerability of Image Processing Components}: Based on the categories of ``DL Model Under Attack'' and ``Attacked Target'', most system-level attacks manipulate the environment in order to affect some image processing components, which are vulnerable to adversarial attacks. \new{It should be noted that incorporating a DL component for image processing requires viewing environmental objects as potential threats and designing defenses accordingly.}
    \item \new{\textbf{Vulnerability of Modules and Systems:} The perception module is the primary target of most attacks, while other modules, such as planning, localization, and control, are rarely targeted directly. While end-to-end AV systems were  expected to be vulnerable to adversarial attacks, as confirmed in the taxonomy, the surprising finding is that MSF systems are also susceptible. However, almost every attack on MSF systems requires white-box knowledge of the DL model and was not designed in a black-box way. We can conclude that, when designing defenses, it is important to consider that the perception module in MSF autonomous vehicles is the most frequently targeted module.} %
    \item \textbf{Attack Propagation}:
    Wang et al.~\cite{wang2023does} highlighted that some existing model-level object evasion attacks in AV fail to produce system-level effects. Therefore, it is crucial to analyze the propagation of these attacks, as the effectiveness of a model-level attack on a system depends on the propagation chain it exploits. During our review of related papers, we identified a lack of research addressing why certain attacks on DL models do not escalate to system-level failures. This gap represents a promising area for further investigation in the field. \new{Filling this gap to understand why an attack leads to system-level failures can help in identifying both attacks and defenses more efficiently and accurately, potentially leading to more secure AVs.}
    \item \textbf{System-level knowledge}: Nearly every attack in the reviewed papers was conducted without any prior knowledge of the system, classifying them as black-box attacks. This demonstrates that effective attacks can still be carried out without access to the source code or detailed information about the autonomous vehicle. \new{Consequently, it should not be assumed that concealing the AV's system information would inherently enhance its security.}
    \item \textbf{Additional Attack Targets}: Existing attacks commonly involve creating adversarial objects or fake road lines to mislead the perception components of an AV. These objects are static and do not move, \new{hence, defense techniques should take into account environmental objects and elements more carefully, as they can play a significant role in enabling or triggering attacks.} However, Hanfeld et al.~\cite{hanfeld2023kidnapping} demonstrated an attack using a moving drone, showing that dynamic objects, such as autonomous or human-controlled units, can also be employed by attackers to carry out their attacks. \new{This points to a potential direction for future work, exploring the use of dynamic objects (e.g., moving vehicles or people) as carriers of adversarial patches.}
\end{enumerate}

\section{Related Work}
\label{sec:related_work}
Biggio \& Roli~\cite{biggio2018wild} proposed a comprehensive survey investigating the vulnerability of ML and the development of suitable countermeasures within the research field of adversarial ML. This work provides an in-depth overview of the evolution of this research area over the past decade and beyond, starting from early investigations into the security of non-DL algorithms to more recent studies aimed at understanding the security properties of DL algorithms in computer vision and cybersecurity. The authors reviewed the main threat models and attacks defined in this context and discussed the significant limitations of current research, as well as the future challenges in designing more secure learning algorithms.

In their subsequent work, Cinà et al.~\cite{cina2023wild} provided a comprehensive survey that systematizes poisoning attacks and defenses in ML over the past 15 years. They begin by categorizing the current threat models and attacks, and then organize the existing defenses accordingly. While their focus is primarily on computer vision applications, they argue that their systematization also encompasses state-of-the-art attacks and defenses for other data modalities. The main difference between these two papers and our work is that their surveys 
cover model-level attacks, while we are interested in system-level attacks.
Our paper aims to provide categorized information for readers interested in understanding how to mount an attack against a whole system, not just one of its DL components in isolation, and how this can lead to system-level failures.

Shen et al.~\cite{shen2024soksemanticaisecurity} systematized the knowledge within the growing area of semantic autonomous driving AI security. They analyzed a total of 53 papers, categorizing them based on research aspects critical to the security field such as the attack/defense targeted AI component, attack/defense goal, attack vector, attack knowledge, defense deployability, defense robustness, and evaluation methodologies. While this paper is the closest to our work, there are remarkable differences. 
First of all the focus, which is \textit{semantic} AI security. As a consequence,
although this paper mentions attacks that can cause system-level failures, its approach to collecting and analyzing papers was not systematically focused on all possible system-level attacks, rather just on those exhibiting a \textit{semantic} gap. Hence, they cover the existing literature on system-level attacks to a quite limited extent. Correspondingly, our taxonomy provides more detailed information about the relationship between each attack and the whole system under test, including how the attacks are performed in the environment, which system components they target, the resulting system-level failures, and the propagation chain that allows an attack to manifest itself. 

\section{Threats to Validity}
\label{sec:threats}
\subsection{Internal Validity}

\textit{Internal validity threats} are caused by the tools and methodology employed to conduct the study. We have identified the following internal validity threats: (1) the query string; (2) the filtering phase; (3) subjectivity of categorization; (4) exclusion criteria.

The \textit{query string} was crafted with a focus on attacks on autonomous vehicles and refined iteratively based on the number of retrieved papers and the presence of pivotal papers in our results. Running the query string from 2017 to 2024 delivered an initial list of approximately 9,700 papers. Although changing the query and pivot papers would alter this initial number, our subsequent steps --- filtering and snowballing --- compensate for this effect. So, we expect that any replications would yield results similar to ours.

In the \textit{filtering phase}, we excluded venues that did not meet our desired quality standards or fell outside the scope of our taxonomy. We also white-listed some high-reputation, related venues to ensure their inclusion, as determining the scope based solely on titles was not feasible. 
While changing the chosen venues might affect the results, manual inspection of the retrieved papers, along with those identified in the snowballing phase, ensured that all included papers were high-value and fully in the scope of our work.

The selection of categories was done by starting from previous model-level surveys and creating new categories when needed. The resulting categories were defined through consensus in weekly meetings held among the authors. The primary task involved comprehensively reading the papers and categorizing them. Consequently, each paper was reviewed by at least two assessors, and weekly meetings were held to resolve any differences in categorization. These measures were taken to ensure accurate categorization and mitigate any \textit{subjectivity threat} possibly affecting the categorization.

We had two main criteria for excluding papers. The first criterion was to exclude attacks not targeting any DL model. The second criterion was to exclude attacks evaluated only at the model level. %
While reviewing the papers, we encountered some that targeted the sensors rather than the DL component, such as those that blinded the sensors. In these cases, the attack could not be attributed to any vulnerable DL model. Thus, these papers were discarded. Some papers claimed real-world evaluation but only involved mounting a camera on a car and taking pictures of traffic signs and surroundings. Since no autonomous vehicle was involved, this merely resulted in creating their own dataset, leading to their exclusion from our taxonomy.
This manual exclusion of papers was carried out through meetings and discussions, and we tried to stick to the two criteria explained above as strictly as possible. However, we acknowledge that other researchers might have different perspectives.

\subsection{External Validity}

\textit{External validity threats} affect the generalizability of our findings beyond the study conducted for this paper.
A replication of this study following our methodology is not ensured to produce exactly the same list of papers, taxonomy categories and taxonomy mapping. However, we have tried to document all the key decisions made in the process, describing the adopted methodology in detail. We have adopted the best practices available for the execution of a systematic literature review~\cite{ISTguidelines}. Hence, we think that our main findings (see our answers to the research questions in Section~\ref{sec:results}) are to a major extent valid beyond our specific instance of the study.

\section{Conclusion}
\label{sec:conclusion}

While model-level adversarial attacks have received huge attention from researchers in the last decades, it is only more recently that researchers considered the possibility of injecting attack vectors into elements of the environment where the system under attack operates, to observe a system-level failure, not just a model-level mis-prediction. The literature on system-level attacks has grown quite rapidly in recent years, going in different directions. Hence, this is the time to organize the existing works into a taxonomy. For this purpose, we retrieved 21 more relevant papers from the literature, and created a taxonomy to categorize them. Our key findings indicate an unbalanced distribution of papers across taxonomy categories; a prevalence of attacks against image processing components \new{and the perception module} in simulation environments; a variety of threat models, which include black-box attacks and white-box attacks, the latter either at inference or at training time; two main \new{attack} chains, consisting of targeted \new{attacks}, aimed at specific system-level failures, and untargeted \new{attacks}.

We also discussed a number of implications for practitioners and researchers. In the former case, they include guidelines and suggestions for security testing (e.g., on risk assessment and attack effectiveness). In the latter case, they indicate possible directions for future work, such as: performing attacks on DL components beyond image processing ones; \new{designing attacks targeting modules outside perception; addressing the vulnerabilities of both multi and single-sensor fusion systems;} investigating the transferability of attacks from a simulator to the real world; propagation of an attack from model to system; system level security testing for the creation of more secure vehicles, including dynamic entities among the environment elements that contribute to a system-level attack.

\section*{Conflict of Interest and Data Availability}

The implementations, source code, data, and experimental results are publicly available at \url{https://zenodo.org/records/16407709}.

\begin{acks}
This work is funded by the European Union's Horizon Europe research and innovation programme under the project Sec4AI4Sec, grant agreement No 101120393.
\end{acks}

\bibliographystyle{IEEEtran}
\balance
\bibliography{onlytaxbib, bib}

\newpage
\appendix
\section{Tables with Paper Mapping} \label{sec:appendix}

Below, we report the full tables with the paper mapping produced by the assessors and agreed upon in the consensus meetings.

\begin{sidewaystable}

\caption{Paper Mapping (Part 1)}
\label{tab:taxonomy1}

\resizebox{0.9\columnwidth}{!}{%
\begin{tabular}{cllll}
\hline
\rowcolor[HTML]{DAE8FC} 
\multicolumn{1}{|c|}{\cellcolor[HTML]{DAE8FC}\textbf{Ref}}      & \multicolumn{1}{c|}{\cellcolor[HTML]{DAE8FC}{\color[HTML]{333333} \textbf{\begin{tabular}[c]{@{}c@{}}App \\ Domain\end{tabular}}}} & \multicolumn{1}{c|}{\cellcolor[HTML]{DAE8FC}{\color[HTML]{333333} \textbf{DL Model Under Attack}}}                                                                         & \multicolumn{1}{c|}{\cellcolor[HTML]{DAE8FC}{\color[HTML]{333333} \textbf{System Under attack}}}                                                                                                 & \multicolumn{1}{c|}{\cellcolor[HTML]{DAE8FC}{\color[HTML]{333333} \textbf{Attack Scenario}}}                                                                                                                                        \\ \hline
\rowcolor[HTML]{FFFFFF} 
\multicolumn{1}{|c|}{\cellcolor[HTML]{FFFFFF}\textbf{{[}1{]}}}  & \multicolumn{1}{l|}{\cellcolor[HTML]{FFFFFF}Cars}                                                                                  & \multicolumn{1}{l|}{\cellcolor[HTML]{FFFFFF}\begin{tabular}[c]{@{}l@{}}SORT(Y5), FairMOT, ByteTrack, BoT-SORT\\ (Perception)\end{tabular}}                                 & \multicolumn{1}{l|}{\cellcolor[HTML]{FFFFFF}Baidu Apollo in LGSVL simulator (MSF)}                                                                                                               & \multicolumn{1}{l|}{\cellcolor[HTML]{FFFFFF}\begin{tabular}[c]{@{}l@{}}AVs changing lanes when another car is in the \\ adjacent lane or approaching an intersection \\ with a stop sign where another car is present\end{tabular}} \\ \hline
\rowcolor[HTML]{EFEFEF} 
\multicolumn{1}{|c|}{\cellcolor[HTML]{EFEFEF}\textbf{{[}2{]}}}  & \multicolumn{1}{l|}{\cellcolor[HTML]{EFEFEF}Drones}                                                                                & \multicolumn{1}{l|}{\cellcolor[HTML]{EFEFEF}A variation of DroNet (E2E)}                                                                                                   & \multicolumn{1}{l|}{\cellcolor[HTML]{EFEFEF}\begin{tabular}[c]{@{}l@{}}Parrot Bebop 2 Drone and \\ Matlab simulation (SSF)\end{tabular}}                                                         & \multicolumn{1}{l|}{\cellcolor[HTML]{EFEFEF}Drones moving to its destination}                                                                                                                                                       \\ \hline
\rowcolor[HTML]{FFFFFF} 
\multicolumn{1}{|c|}{\cellcolor[HTML]{FFFFFF}\textbf{{[}3{]}}}  & \multicolumn{1}{l|}{\cellcolor[HTML]{FFFFFF}Cars}                                                                                  & \multicolumn{1}{l|}{\cellcolor[HTML]{FFFFFF}\begin{tabular}[c]{@{}l@{}}OpenPilot's Adaptive Cruise\\ Control ACC system (Perception)\end{tabular}}                         & \multicolumn{1}{l|}{\cellcolor[HTML]{FFFFFF}CARLA and Baidu Apollo (MSF)}                                                                                                                        & \multicolumn{1}{l|}{\cellcolor[HTML]{FFFFFF}AV driving behind the attacker's vehicle}                                                                                                                                               \\ \hline
\rowcolor[HTML]{EFEFEF} 
\multicolumn{1}{|c|}{\cellcolor[HTML]{EFEFEF}\textbf{{[}4{]}}}  & \multicolumn{1}{l|}{\cellcolor[HTML]{EFEFEF}\begin{tabular}[c]{@{}l@{}}Cars, \\ Drones\end{tabular}}                               & \multicolumn{1}{l|}{\cellcolor[HTML]{EFEFEF}YOLO v5 (Perception)}                                                                                                          & \multicolumn{1}{l|}{\cellcolor[HTML]{EFEFEF}\begin{tabular}[c]{@{}l@{}}Vision-based guidance system in \\ CARLA or AirSim (SSF)\end{tabular}}                                                    & \multicolumn{1}{l|}{\cellcolor[HTML]{EFEFEF}AVs following their target}                                                                                                                                                             \\ \hline
\rowcolor[HTML]{FFFFFF} 
\multicolumn{1}{|c|}{\cellcolor[HTML]{FFFFFF}\textbf{{[}5{]}}}  & \multicolumn{1}{l|}{\cellcolor[HTML]{FFFFFF}Cars}                                                                                  & \multicolumn{1}{l|}{\cellcolor[HTML]{FFFFFF}Dave2 (E2E)}                                                                                                                   & \multicolumn{1}{l|}{\cellcolor[HTML]{FFFFFF}\begin{tabular}[c]{@{}l@{}}An AV vehicle (hopper vehicle) in BeamNG\\ (SSF)\end{tabular}}                                                            & \multicolumn{1}{l|}{\cellcolor[HTML]{FFFFFF}AVs driving past a roadside billboard.}                                                                                                                                                 \\ \hline
\rowcolor[HTML]{EFEFEF} 
\multicolumn{1}{|c|}{\cellcolor[HTML]{EFEFEF}\textbf{{[}6{]}}}  & \multicolumn{1}{l|}{\cellcolor[HTML]{EFEFEF}Drones}                                                                                & \multicolumn{1}{l|}{\cellcolor[HTML]{EFEFEF}PULP Frontnet (Perception)}                                                                                                    & \multicolumn{1}{l|}{\cellcolor[HTML]{EFEFEF}\begin{tabular}[c]{@{}l@{}}A human-following drone control \\ system known as the nano multirotor, \\ the Crazyflie by Bitcraze. (SSF)\end{tabular}} & \multicolumn{1}{l|}{\cellcolor[HTML]{EFEFEF}Autonomous drones following a human target}                                                                                                                                             \\ \hline
\rowcolor[HTML]{FFFFFF} 
\multicolumn{1}{|c|}{\cellcolor[HTML]{FFFFFF}\textbf{{[}7{]}}}  & \multicolumn{1}{l|}{\cellcolor[HTML]{FFFFFF}Cars}                                                                                  & \multicolumn{1}{l|}{\cellcolor[HTML]{FFFFFF}YOLO v2, v3, v5 (Perception)}                                                                                                  & \multicolumn{1}{l|}{\cellcolor[HTML]{FFFFFF}Baidu Apollo in LGSVL simulator (MSF)}                                                                                                               & \multicolumn{1}{l|}{\cellcolor[HTML]{FFFFFF}\begin{tabular}[c]{@{}l@{}}AVs approaching a critical physical road object on \\ a sunny day, such as a stop sign or pedestrian\end{tabular}}                                           \\ \hline
\rowcolor[HTML]{EFEFEF} 
\multicolumn{1}{|c|}{\cellcolor[HTML]{EFEFEF}\textbf{{[}8{]}}}  & \multicolumn{1}{l|}{\cellcolor[HTML]{EFEFEF}Cars}                                                                                  & \multicolumn{1}{l|}{\cellcolor[HTML]{EFEFEF}\begin{tabular}[c]{@{}l@{}}PointPillars, VoxelNet, V2VNet, CoBEVT, \\ FPV-RCNN (Perception)\end{tabular}}                      & \multicolumn{1}{l|}{\cellcolor[HTML]{EFEFEF}\begin{tabular}[c]{@{}l@{}}AVs utilizing LiDAR or GPS inform-\\ ation from other cars in Baidu Apollo (MSF)\end{tabular}}                            & \multicolumn{1}{l|}{\cellcolor[HTML]{EFEFEF}\begin{tabular}[c]{@{}l@{}}Multiple AVs driving and jointly performing \\ collaborative perception tasks\end{tabular}}                                                                  \\ \hline
\rowcolor[HTML]{FFFFFF} 
\multicolumn{1}{|c|}{\cellcolor[HTML]{FFFFFF}\textbf{{[}9{]}}}  & \multicolumn{1}{l|}{\cellcolor[HTML]{FFFFFF}Cars}                                                                                  & \multicolumn{1}{l|}{\cellcolor[HTML]{FFFFFF}Nexar \&amp; YOLO v4 (Perception)}                                                                                             & \multicolumn{1}{l|}{\cellcolor[HTML]{FFFFFF}Baidu Apollo in LGSVL simulator (MSF)}                                                                                                               & \multicolumn{1}{l|}{\cellcolor[HTML]{FFFFFF}AVs approaching a traffic light}                                                                                                                                                        \\ \hline
\rowcolor[HTML]{EFEFEF} 
\multicolumn{1}{|c|}{\cellcolor[HTML]{EFEFEF}\textbf{{[}10{]}}} & \multicolumn{1}{l|}{\cellcolor[HTML]{EFEFEF}Cars}                                                                                  & \multicolumn{1}{l|}{\cellcolor[HTML]{EFEFEF}\begin{tabular}[c]{@{}l@{}}DRL in decision-making\\ module of SUMO (Planning)\end{tabular}}                                    & \multicolumn{1}{l|}{\cellcolor[HTML]{EFEFEF}\begin{tabular}[c]{@{}l@{}}Microscopic traffic simulator SUMO \\ And intelligent driver model (SSF)\end{tabular}}                                    & \multicolumn{1}{l|}{\cellcolor[HTML]{EFEFEF}\begin{tabular}[c]{@{}l@{}}AVs driving in one or two lanes, following the \\ malicious leading vehicle in traffic\end{tabular}}                                                         \\ \hline
\rowcolor[HTML]{FFFFFF} 
\multicolumn{1}{|c|}{\cellcolor[HTML]{FFFFFF}\textbf{{[}11{]}}} & \multicolumn{1}{l|}{\cellcolor[HTML]{FFFFFF}Cars}                                                                                  & \multicolumn{1}{l|}{\cellcolor[HTML]{FFFFFF}ResNet-34 (E2E)}                                                                                                               & \multicolumn{1}{l|}{\cellcolor[HTML]{FFFFFF}Learning by cheating agent in CARLA (SSF)}                                                                                                           & \multicolumn{1}{l|}{\cellcolor[HTML]{FFFFFF}AVs on a road with pedestrians and other vehicles}                                                                                                                                      \\ \hline
\rowcolor[HTML]{EFEFEF} 
\multicolumn{1}{|c|}{\cellcolor[HTML]{EFEFEF}\textbf{{[}12{]}}} & \multicolumn{1}{l|}{\cellcolor[HTML]{EFEFEF}Cars}                                                                                  & \multicolumn{1}{l|}{\cellcolor[HTML]{EFEFEF}\begin{tabular}[c]{@{}l@{}}OpenPilot's Automated Lane Centering \\ (ALC) (E2E)\end{tabular}}                                   & \multicolumn{1}{l|}{\cellcolor[HTML]{EFEFEF}\begin{tabular}[c]{@{}l@{}}Automated Lane Centering (ALC) \\ system in LGSVL (SSF)\end{tabular}}                                                     & \multicolumn{1}{l|}{\cellcolor[HTML]{EFEFEF}AVs driving}                                                                                                                                                                            \\ \hline
\rowcolor[HTML]{FFFFFF} 
\multicolumn{1}{|c|}{\cellcolor[HTML]{FFFFFF}\textbf{{[}13{]}}} & \multicolumn{1}{l|}{\cellcolor[HTML]{FFFFFF}Cars}                                                                                  & \multicolumn{1}{l|}{\cellcolor[HTML]{FFFFFF}\begin{tabular}[c]{@{}l@{}}(A5-L, A5-C) In-road obstacle detection \\ (Perception)\end{tabular}}                               & \multicolumn{1}{l|}{\cellcolor[HTML]{FFFFFF}Baidu Apollo in LGSVL simulator v5 (MSF)}                                                                                                            & \multicolumn{1}{l|}{\cellcolor[HTML]{FFFFFF}Av on a single lane road}                                                                                                                                                               \\ \hline
\rowcolor[HTML]{EFEFEF} 
\multicolumn{1}{|c|}{\cellcolor[HTML]{EFEFEF}\textbf{{[}14{]}}} & \multicolumn{1}{l|}{\cellcolor[HTML]{EFEFEF}Cars}                                                                                  & \multicolumn{1}{l|}{\cellcolor[HTML]{EFEFEF}\begin{tabular}[c]{@{}l@{}}Tesla's APE lane recognition (AutoPilot \\ ECU, Electronic Control Unit) (Perception)\end{tabular}} & \multicolumn{1}{l|}{\cellcolor[HTML]{EFEFEF}Tesla model S (SSF)}                                                                                                                                 & \multicolumn{1}{l|}{\cellcolor[HTML]{EFEFEF}\begin{tabular}[c]{@{}l@{}}AVs on roads driving into empty spaces \\ in road line markings\end{tabular}}                                                                                \\ \hline
\rowcolor[HTML]{FFFFFF} 
\multicolumn{1}{|c|}{\cellcolor[HTML]{FFFFFF}\textbf{{[}15{]}}} & \multicolumn{1}{l|}{\cellcolor[HTML]{FFFFFF}Cars}                                                                                  & \multicolumn{1}{l|}{\cellcolor[HTML]{FFFFFF}\begin{tabular}[c]{@{}l@{}}PointRCNN and PointPillar and PV-RCNN \\ (Perception)\end{tabular}}                                 & \multicolumn{1}{l|}{\cellcolor[HTML]{FFFFFF}\begin{tabular}[c]{@{}l@{}}Baidu Apollo in LGSVL simulator \\ (Lincoln MKZ car) (MSF)\end{tabular}}                                                  & \multicolumn{1}{l|}{\cellcolor[HTML]{FFFFFF}AVs driving on a single lane road}                                                                                                                                                      \\ \hline
\rowcolor[HTML]{EFEFEF} 
\multicolumn{1}{|c|}{\cellcolor[HTML]{EFEFEF}\textbf{{[}16{]}}} & \multicolumn{1}{l|}{\cellcolor[HTML]{EFEFEF}Cars}                                                                                  & \multicolumn{1}{l|}{\cellcolor[HTML]{EFEFEF}Kalman Filter in YOLO v3 (Perception)}                                                                                         & \multicolumn{1}{l|}{\cellcolor[HTML]{EFEFEF}Baidu Apollo in LGSVL simulator (MSF)}                                                                                                               & \multicolumn{1}{l|}{\cellcolor[HTML]{EFEFEF}\begin{tabular}[c]{@{}l@{}}AVs on a road with pedestrians and vehicles in \\ front of the car\end{tabular}}                                                                             \\ \hline
\rowcolor[HTML]{FFFFFF} 
\multicolumn{1}{|c|}{\cellcolor[HTML]{FFFFFF}\textbf{{[}17{]}}} & \multicolumn{1}{l|}{\cellcolor[HTML]{FFFFFF}Cars}                                                                                  & \multicolumn{1}{l|}{\cellcolor[HTML]{FFFFFF}\begin{tabular}[c]{@{}l@{}}Conditional Imitation and Reinforcement\\ learning (E2E) in CARLA\end{tabular}}                     & \multicolumn{1}{l|}{\cellcolor[HTML]{FFFFFF}End to end driving system in CARLA (SSF)}                                                                                                            & \multicolumn{1}{l|}{\cellcolor[HTML]{FFFFFF}AVs in turns and intersections}                                                                                                                                                         \\ \hline
\rowcolor[HTML]{EFEFEF} 
\multicolumn{1}{|c|}{\cellcolor[HTML]{EFEFEF}\textbf{{[}18{]}}} & \multicolumn{1}{l|}{\cellcolor[HTML]{EFEFEF}Cars}                                                                                  & \multicolumn{1}{l|}{\cellcolor[HTML]{EFEFEF}DriveNet (an extension of Dave2) (E2E)}                                                                                        & \multicolumn{1}{l|}{\cellcolor[HTML]{EFEFEF}Autonomous Driving agent in CARLA (SSF)}                                                                                                             & \multicolumn{1}{l|}{\cellcolor[HTML]{EFEFEF}AVs driving past a roadside billboard}                                                                                                                                                  \\ \hline
\rowcolor[HTML]{FFFFFF} 
\multicolumn{1}{|c|}{\cellcolor[HTML]{FFFFFF}\textbf{{[}19{]}}} & \multicolumn{1}{l|}{\cellcolor[HTML]{FFFFFF}Cars}                                                                                  & \multicolumn{1}{l|}{\cellcolor[HTML]{FFFFFF}\begin{tabular}[c]{@{}l@{}}Faster\_rcnn\_inception\_v2 and Tesla's \\ (Perception) model\end{tabular}}                         & \multicolumn{1}{l|}{\cellcolor[HTML]{FFFFFF}\begin{tabular}[c]{@{}l@{}}Tesla Model X HW 2.5/3 \&amp;amp; Renault \\ Captur (equipped with Mobileye 630) (MSF)\end{tabular}}                      & \multicolumn{1}{l|}{\cellcolor[HTML]{FFFFFF}AVs driving}                                                                                                                                                                            \\ \hline
\rowcolor[HTML]{EFEFEF} 
\multicolumn{1}{|c|}{\cellcolor[HTML]{EFEFEF}\textbf{{[}20{]}}} & \multicolumn{1}{l|}{\cellcolor[HTML]{EFEFEF}Cars}                                                                                  & \multicolumn{1}{l|}{\cellcolor[HTML]{EFEFEF}\begin{tabular}[c]{@{}l@{}}DNN of the Lidar-based (Perception) \\ module in Baidu Apollo\end{tabular}}                         & \multicolumn{1}{l|}{\cellcolor[HTML]{EFEFEF}Driving agent in Baidu Apollo (MSF)}                                                                                                                 & \multicolumn{1}{l|}{\cellcolor[HTML]{EFEFEF}AVs driving}                                                                                                                                                                            \\ \hline
\rowcolor[HTML]{FFFFFF} 
\multicolumn{1}{|c|}{\cellcolor[HTML]{FFFFFF}\textbf{{[}21{]}}} & \multicolumn{1}{l|}{\cellcolor[HTML]{FFFFFF}Cars}                                                                                  & \multicolumn{1}{l|}{\cellcolor[HTML]{FFFFFF}NN model of Udacity Simulator (E2E)}                                                                                           & \multicolumn{1}{l|}{\cellcolor[HTML]{FFFFFF}Udacity Simulator (SSF)}                                                                                                                             & \multicolumn{1}{l|}{\cellcolor[HTML]{FFFFFF}AVs driving}                                                                                                                                                                            \\ \hline
\multicolumn{5}{c}{*MSF stands for Multi-Sensor Fusion and SSF stands for Single-Sensor Fusion}                                                                                                                                                                                                                                                                                                                                                                                                                                                                                                                                                                                                                                                                                                                           
\end{tabular}
}

\end{sidewaystable}

\begin{sidewaystable}
\caption{Paper Mapping (Part 2)}
\label{tab:taxonomy2}

\resizebox{0.85\columnwidth}{!}{%
\begin{tabular}{cllllll}
\hline
\rowcolor[HTML]{DAE8FC} 
\multicolumn{1}{|c|}{\cellcolor[HTML]{DAE8FC}\textbf{Ref}}      & \multicolumn{1}{c|}{\cellcolor[HTML]{DAE8FC}{\color[HTML]{333333} \textbf{Attacked Target}}}                                                      & \multicolumn{1}{c|}{\cellcolor[HTML]{DAE8FC}{\color[HTML]{333333} \textbf{Attacker's Capability}}}                                                                                            & \multicolumn{1}{c|}{\cellcolor[HTML]{DAE8FC}{\color[HTML]{333333} \textbf{\begin{tabular}[c]{@{}c@{}}Attack \\ Strategy\end{tabular}}}} & \multicolumn{1}{c|}{\cellcolor[HTML]{DAE8FC}{\color[HTML]{333333} \textbf{\begin{tabular}[c]{@{}c@{}}Attacker's \\ System/Model \\ knowledge\end{tabular}}}} & \multicolumn{1}{c|}{\cellcolor[HTML]{DAE8FC}{\color[HTML]{333333} \textbf{\begin{tabular}[c]{@{}c@{}}Attack/Error \\ Specificity\end{tabular}}}} & \multicolumn{1}{c|}{\cellcolor[HTML]{DAE8FC}{\color[HTML]{333333} \textbf{\begin{tabular}[c]{@{}c@{}}Failure \\ Specificity\end{tabular}}}} \\ \hline
\rowcolor[HTML]{FFFFFF} 
\multicolumn{1}{|c|}{\cellcolor[HTML]{FFFFFF}\textbf{{[}1{]}}}  & \multicolumn{1}{l|}{\cellcolor[HTML]{FFFFFF}Input image}                                                                                          & \multicolumn{1}{l|}{\cellcolor[HTML]{FFFFFF}\begin{tabular}[c]{@{}l@{}}Purturbing the image input to insert fake \\ bounding boxes into it\end{tabular}}                                      & \multicolumn{1}{l|}{\cellcolor[HTML]{FFFFFF}Evasion}                                                                                    & \multicolumn{1}{l|}{\cellcolor[HTML]{FFFFFF}WM / BS *}                                                                                                       & \multicolumn{1}{l|}{\cellcolor[HTML]{FFFFFF}AG / EG *}                                                                                           & \multicolumn{1}{l|}{\cellcolor[HTML]{FFFFFF}Generic}                                                                                        \\ \hline
\rowcolor[HTML]{EFEFEF} 
\multicolumn{1}{|c|}{\cellcolor[HTML]{EFEFEF}\textbf{{[}2{]}}}  & \multicolumn{1}{l|}{\cellcolor[HTML]{EFEFEF}\begin{tabular}[c]{@{}l@{}}Environment (anything that can \\ hold a patch horizontally)\end{tabular}} & \multicolumn{1}{l|}{\cellcolor[HTML]{EFEFEF}\begin{tabular}[c]{@{}l@{}}Attaching the patch to anything that can \\ hold it horizontally, facing the drone.\end{tabular}}                      & \multicolumn{1}{l|}{\cellcolor[HTML]{EFEFEF}Evasion}                                                                                    & \multicolumn{1}{l|}{\cellcolor[HTML]{EFEFEF}WM / GS}                                                                                                         & \multicolumn{1}{l|}{\cellcolor[HTML]{EFEFEF}AG / ES}                                                                                             & \multicolumn{1}{l|}{\cellcolor[HTML]{EFEFEF}Specific}                                                                                       \\ \hline
\rowcolor[HTML]{FFFFFF} 
\multicolumn{1}{|c|}{\cellcolor[HTML]{FFFFFF}\textbf{{[}3{]}}}  & \multicolumn{1}{l|}{\cellcolor[HTML]{FFFFFF}Vehicles and Trucks}                                                                                  & \multicolumn{1}{l|}{\cellcolor[HTML]{FFFFFF}\begin{tabular}[c]{@{}l@{}}Placing the patch on the back of a vehicle\\ and drive it in front of the AV\end{tabular}}                             & \multicolumn{1}{l|}{\cellcolor[HTML]{FFFFFF}Evasion}                                                                                    & \multicolumn{1}{l|}{\cellcolor[HTML]{FFFFFF}WM / BS}                                                                                                         & \multicolumn{1}{l|}{\cellcolor[HTML]{FFFFFF}AS / ES}                                                                                             & \multicolumn{1}{l|}{\cellcolor[HTML]{FFFFFF}Specific}                                                                                       \\ \hline
\rowcolor[HTML]{EFEFEF} 
\multicolumn{1}{|c|}{\cellcolor[HTML]{EFEFEF}\textbf{{[}4{]}}}  & \multicolumn{1}{l|}{\cellcolor[HTML]{EFEFEF}Input image}                                                                                          & \multicolumn{1}{l|}{\cellcolor[HTML]{EFEFEF}\begin{tabular}[c]{@{}l@{}}Inserting the malware into the system \\ and training the perturbations on the \\ real training set\end{tabular}}      & \multicolumn{1}{l|}{\cellcolor[HTML]{EFEFEF}Evasion}                                                                                    & \multicolumn{1}{l|}{\cellcolor[HTML]{EFEFEF}WM / BS}                                                                                                         & \multicolumn{1}{l|}{\cellcolor[HTML]{EFEFEF}AG / EG}                                                                                             & \multicolumn{1}{l|}{\cellcolor[HTML]{EFEFEF}Generic}                                                                                        \\ \hline
\rowcolor[HTML]{FFFFFF} 
\multicolumn{1}{|c|}{\cellcolor[HTML]{FFFFFF}\textbf{{[}5{]}}}  & \multicolumn{1}{l|}{\cellcolor[HTML]{FFFFFF}\begin{tabular}[c]{@{}l@{}}Billboards (placing their \\ own malicious billboard)\end{tabular}}        & \multicolumn{1}{l|}{\cellcolor[HTML]{FFFFFF}\begin{tabular}[c]{@{}l@{}}Installing a perturbed billboard or altering \\ an existing one\end{tabular}}                                          & \multicolumn{1}{l|}{\cellcolor[HTML]{FFFFFF}Evasion}                                                                                    & \multicolumn{1}{l|}{\cellcolor[HTML]{FFFFFF}WM / BS}                                                                                                         & \multicolumn{1}{l|}{\cellcolor[HTML]{FFFFFF}AS / EG}                                                                                             & \multicolumn{1}{l|}{\cellcolor[HTML]{FFFFFF}Generic}                                                                                        \\ \hline
\rowcolor[HTML]{EFEFEF} 
\multicolumn{1}{|c|}{\cellcolor[HTML]{EFEFEF}\textbf{{[}6{]}}}  & \multicolumn{1}{l|}{\cellcolor[HTML]{EFEFEF}Input image}                                                                                          & \multicolumn{1}{l|}{\cellcolor[HTML]{EFEFEF}Moving a printed patch in front of a drone}                                                                                                       & \multicolumn{1}{l|}{\cellcolor[HTML]{EFEFEF}Evasion}                                                                                    & \multicolumn{1}{l|}{\cellcolor[HTML]{EFEFEF}WM / BS}                                                                                                         & \multicolumn{1}{l|}{\cellcolor[HTML]{EFEFEF}AG / ES}                                                                                             & \multicolumn{1}{l|}{\cellcolor[HTML]{EFEFEF}Specific}                                                                                       \\ \hline
\rowcolor[HTML]{FFFFFF} 
\multicolumn{1}{|c|}{\cellcolor[HTML]{FFFFFF}\textbf{{[}7{]}}}  & \multicolumn{1}{l|}{\cellcolor[HTML]{FFFFFF}Stop signs and pedestrians}                                                                           & \multicolumn{1}{l|}{\cellcolor[HTML]{FFFFFF}\begin{tabular}[c]{@{}l@{}}Attaching adversarial patches on stop \\ signs or pedestrians' shirts.\end{tabular}}                                   & \multicolumn{1}{l|}{\cellcolor[HTML]{FFFFFF}Evasion}                                                                                    & \multicolumn{1}{l|}{\cellcolor[HTML]{FFFFFF}WM / BS}                                                                                                         & \multicolumn{1}{l|}{\cellcolor[HTML]{FFFFFF}AS / ES}                                                                                             & \multicolumn{1}{l|}{\cellcolor[HTML]{FFFFFF}Specific}                                                                                       \\ \hline
\rowcolor[HTML]{EFEFEF} 
\multicolumn{1}{|c|}{\cellcolor[HTML]{EFEFEF}\textbf{{[}8{]}}}  & \multicolumn{1}{l|}{\cellcolor[HTML]{EFEFEF}Objects in the shared feature map}                                                                    & \multicolumn{1}{l|}{\cellcolor[HTML]{EFEFEF}\begin{tabular}[c]{@{}l@{}}Manipulating shared LiDAR vision data \\ by spoofing or removing objects from it\end{tabular}}                         & \multicolumn{1}{l|}{\cellcolor[HTML]{EFEFEF}Evasion}                                                                                    & \multicolumn{1}{l|}{\cellcolor[HTML]{EFEFEF}\begin{tabular}[c]{@{}l@{}}WM \&amp;amp; BM\\ / GS\end{tabular}}                                                 & \multicolumn{1}{l|}{\cellcolor[HTML]{EFEFEF}AS / ES}                                                                                             & \multicolumn{1}{l|}{\cellcolor[HTML]{EFEFEF}Specific}                                                                                       \\ \hline
\rowcolor[HTML]{FFFFFF} 
\multicolumn{1}{|c|}{\cellcolor[HTML]{FFFFFF}\textbf{{[}9{]}}}  & \multicolumn{1}{l|}{\cellcolor[HTML]{FFFFFF}Camera}                                                                                               & \multicolumn{1}{l|}{\cellcolor[HTML]{FFFFFF}\begin{tabular}[c]{@{}l@{}}Directing a laser beam towards the \\ camera of the victim\end{tabular}}                                               & \multicolumn{1}{l|}{\cellcolor[HTML]{FFFFFF}Evasion}                                                                                    & \multicolumn{1}{l|}{\cellcolor[HTML]{FFFFFF}WM / BS}                                                                                                         & \multicolumn{1}{l|}{\cellcolor[HTML]{FFFFFF}AS / ES}                                                                                             & \multicolumn{1}{l|}{\cellcolor[HTML]{FFFFFF}Specific}                                                                                       \\ \hline
\rowcolor[HTML]{EFEFEF} 
\multicolumn{1}{|c|}{\cellcolor[HTML]{EFEFEF}\textbf{{[}10{]}}} & \multicolumn{1}{l|}{\cellcolor[HTML]{EFEFEF}Training data}                                                                                        & \multicolumn{1}{l|}{\cellcolor[HTML]{EFEFEF}\begin{tabular}[c]{@{}l@{}}Injecting a backdoor into the model by \\ retraining it with poisoned data.\end{tabular}}                              & \multicolumn{1}{l|}{\cellcolor[HTML]{EFEFEF}Poisoning}                                                                                  & \multicolumn{1}{l|}{\cellcolor[HTML]{EFEFEF}WM / BS}                                                                                                         & \multicolumn{1}{l|}{\cellcolor[HTML]{EFEFEF}AS / ES}                                                                                             & \multicolumn{1}{l|}{\cellcolor[HTML]{EFEFEF}Specific}                                                                                       \\ \hline
\rowcolor[HTML]{FFFFFF} 
\multicolumn{1}{|c|}{\cellcolor[HTML]{FFFFFF}\textbf{{[}11{]}}} & \multicolumn{1}{l|}{\cellcolor[HTML]{FFFFFF}Input image}                                                                                          & \multicolumn{1}{l|}{\cellcolor[HTML]{FFFFFF}\begin{tabular}[c]{@{}l@{}}Manipulating the neurons and weights \\ of the NN, or manipulating input images\end{tabular}}                          & \multicolumn{1}{l|}{\cellcolor[HTML]{FFFFFF}Evasion}                                                                                    & \multicolumn{1}{l|}{\cellcolor[HTML]{FFFFFF}\begin{tabular}[c]{@{}l@{}}WM \&amp;amp; BM \\ / BS\end{tabular}}                                                & \multicolumn{1}{l|}{\cellcolor[HTML]{FFFFFF}AG / EG}                                                                                             & \multicolumn{1}{l|}{\cellcolor[HTML]{FFFFFF}Specific}                                                                                       \\ \hline
\rowcolor[HTML]{EFEFEF} 
\multicolumn{1}{|c|}{\cellcolor[HTML]{EFEFEF}\textbf{{[}12{]}}} & \multicolumn{1}{l|}{\cellcolor[HTML]{EFEFEF}Roads}                                                                                                & \multicolumn{1}{l|}{\cellcolor[HTML]{EFEFEF}Placing patches on the road surface}                                                                                                              & \multicolumn{1}{l|}{\cellcolor[HTML]{EFEFEF}Evasion}                                                                                    & \multicolumn{1}{l|}{\cellcolor[HTML]{EFEFEF}WM / BS}                                                                                                         & \multicolumn{1}{l|}{\cellcolor[HTML]{EFEFEF}AG / ES}                                                                                             & \multicolumn{1}{l|}{\cellcolor[HTML]{EFEFEF}Specific}                                                                                       \\ \hline
\rowcolor[HTML]{FFFFFF} 
\multicolumn{1}{|c|}{\cellcolor[HTML]{FFFFFF}\textbf{{[}13{]}}} & \multicolumn{1}{l|}{\cellcolor[HTML]{FFFFFF}Roads}                                                                                                & \multicolumn{1}{l|}{\cellcolor[HTML]{FFFFFF}\begin{tabular}[c]{@{}l@{}}Placing a 3D traffic cone that resembles \\ a broken cone on the road\end{tabular}}                                    & \multicolumn{1}{l|}{\cellcolor[HTML]{FFFFFF}Evasion}                                                                                    & \multicolumn{1}{l|}{\cellcolor[HTML]{FFFFFF}WM / BS}                                                                                                         & \multicolumn{1}{l|}{\cellcolor[HTML]{FFFFFF}AS / ES}                                                                                             & \multicolumn{1}{l|}{\cellcolor[HTML]{FFFFFF}Specific}                                                                                       \\ \hline
\rowcolor[HTML]{EFEFEF} 
\multicolumn{1}{|c|}{\cellcolor[HTML]{EFEFEF}\textbf{{[}14{]}}} & \multicolumn{1}{l|}{\cellcolor[HTML]{EFEFEF}Roads}                                                                                                & \multicolumn{1}{l|}{\cellcolor[HTML]{EFEFEF}\begin{tabular}[c]{@{}l@{}}Painting or attaching perturbations on \\ the road surface\end{tabular}}                                               & \multicolumn{1}{l|}{\cellcolor[HTML]{EFEFEF}Evasion}                                                                                    & \multicolumn{1}{l|}{\cellcolor[HTML]{EFEFEF}BM / BS}                                                                                                         & \multicolumn{1}{l|}{\cellcolor[HTML]{EFEFEF}AG / ES}                                                                                             & \multicolumn{1}{l|}{\cellcolor[HTML]{EFEFEF}Specific}                                                                                       \\ \hline
\rowcolor[HTML]{FFFFFF} 
\multicolumn{1}{|c|}{\cellcolor[HTML]{FFFFFF}\textbf{{[}15{]}}} & \multicolumn{1}{l|}{\cellcolor[HTML]{FFFFFF}Roadsides}                                                                                            & \multicolumn{1}{l|}{\cellcolor[HTML]{FFFFFF}Put adversarial objects on the roadside}                                                                                                          & \multicolumn{1}{l|}{\cellcolor[HTML]{FFFFFF}Evasion}                                                                                    & \multicolumn{1}{l|}{\cellcolor[HTML]{FFFFFF}\begin{tabular}[c]{@{}l@{}}WM \&amp;amp; BM \\ / BS\end{tabular}}                                                & \multicolumn{1}{l|}{\cellcolor[HTML]{FFFFFF}AG / ES}                                                                                             & \multicolumn{1}{l|}{\cellcolor[HTML]{FFFFFF}Generic}                                                                                        \\ \hline
\rowcolor[HTML]{EFEFEF} 
\multicolumn{1}{|c|}{\cellcolor[HTML]{EFEFEF}\textbf{{[}16{]}}} & \multicolumn{1}{l|}{\cellcolor[HTML]{EFEFEF}Input image}                                                                                          & \multicolumn{1}{l|}{\cellcolor[HTML]{EFEFEF}\begin{tabular}[c]{@{}l@{}}Installing malware to gain access to the \\ live camera feed and modify its ourput\end{tabular}}                       & \multicolumn{1}{l|}{\cellcolor[HTML]{EFEFEF}Evasion}                                                                                    & \multicolumn{1}{l|}{\cellcolor[HTML]{EFEFEF}WM / WS}                                                                                                         & \multicolumn{1}{l|}{\cellcolor[HTML]{EFEFEF}AS / ES}                                                                                             & \multicolumn{1}{l|}{\cellcolor[HTML]{EFEFEF}Specific}                                                                                       \\ \hline
\rowcolor[HTML]{FFFFFF} 
\multicolumn{1}{|c|}{\cellcolor[HTML]{FFFFFF}\textbf{{[}17{]}}} & \multicolumn{1}{l|}{\cellcolor[HTML]{FFFFFF}Roads}                                                                                                & \multicolumn{1}{l|}{\cellcolor[HTML]{FFFFFF}\begin{tabular}[c]{@{}l@{}}Attaching stickers or painting adversarial \\ lane markings on the road.\end{tabular}}                                 & \multicolumn{1}{l|}{\cellcolor[HTML]{FFFFFF}Evasion}                                                                                    & \multicolumn{1}{l|}{\cellcolor[HTML]{FFFFFF}BM / BS}                                                                                                         & \multicolumn{1}{l|}{\cellcolor[HTML]{FFFFFF}AS / ES}                                                                                             & \multicolumn{1}{l|}{\cellcolor[HTML]{FFFFFF}Specific}                                                                                       \\ \hline
\rowcolor[HTML]{EFEFEF} 
\multicolumn{1}{|c|}{\cellcolor[HTML]{EFEFEF}\textbf{{[}18{]}}} & \multicolumn{1}{l|}{\cellcolor[HTML]{EFEFEF}Billboard}                                                                                            & \multicolumn{1}{l|}{\cellcolor[HTML]{EFEFEF}\begin{tabular}[c]{@{}l@{}}Placing adversarial patches on billboards \\ located on the right side of the roadside\end{tabular}}                   & \multicolumn{1}{l|}{\cellcolor[HTML]{EFEFEF}Evasion}                                                                                    & \multicolumn{1}{l|}{\cellcolor[HTML]{EFEFEF}WM / BS}                                                                                                         & \multicolumn{1}{l|}{\cellcolor[HTML]{EFEFEF}AS / EG}                                                                                             & \multicolumn{1}{l|}{\cellcolor[HTML]{EFEFEF}Specific}                                                                                       \\ \hline
\rowcolor[HTML]{FFFFFF} 
\multicolumn{1}{|c|}{\cellcolor[HTML]{FFFFFF}\textbf{{[}19{]}}} & \multicolumn{1}{l|}{\cellcolor[HTML]{FFFFFF}\begin{tabular}[c]{@{}l@{}}The environment (anywhere \\ with a smooth surface)\end{tabular}}          & \multicolumn{1}{l|}{\cellcolor[HTML]{FFFFFF}\begin{tabular}[c]{@{}l@{}}Projecting a phantom using a projector \\ or embedding the phantom on a digital \\ advertising billboard\end{tabular}} & \multicolumn{1}{l|}{\cellcolor[HTML]{FFFFFF}Evasion}                                                                                    & \multicolumn{1}{l|}{\cellcolor[HTML]{FFFFFF}BM / BS}                                                                                                         & \multicolumn{1}{l|}{\cellcolor[HTML]{FFFFFF}AG / EG}                                                                                             & \multicolumn{1}{l|}{\cellcolor[HTML]{FFFFFF}Generic}                                                                                        \\ \hline
\rowcolor[HTML]{EFEFEF} 
\multicolumn{1}{|c|}{\cellcolor[HTML]{EFEFEF}\textbf{{[}20{]}}} & \multicolumn{1}{l|}{\cellcolor[HTML]{EFEFEF}LIDAR Sensor}                                                                                         & \multicolumn{1}{l|}{\cellcolor[HTML]{EFEFEF}\begin{tabular}[c]{@{}l@{}}Injecting spoofed LiDAR data points \\ by shooting lasers at the AV\end{tabular}}                                      & \multicolumn{1}{l|}{\cellcolor[HTML]{EFEFEF}Evasion}                                                                                    & \multicolumn{1}{l|}{\cellcolor[HTML]{EFEFEF}WM / BS}                                                                                                         & \multicolumn{1}{l|}{\cellcolor[HTML]{EFEFEF}AG / ES}                                                                                             & \multicolumn{1}{l|}{\cellcolor[HTML]{EFEFEF}Specific}                                                                                       \\ \hline
\rowcolor[HTML]{FFFFFF} 
\multicolumn{1}{|c|}{\cellcolor[HTML]{FFFFFF}\textbf{{[}21{]}}} & \multicolumn{1}{l|}{\cellcolor[HTML]{FFFFFF}Billboards}                                                                                           & \multicolumn{1}{l|}{\cellcolor[HTML]{FFFFFF}\begin{tabular}[c]{@{}l@{}}Retraining the model with Trojan data and \\ attaching it to billboards.\end{tabular}}                                 & \multicolumn{1}{l|}{\cellcolor[HTML]{FFFFFF}Poisoning}                                                                                  & \multicolumn{1}{l|}{\cellcolor[HTML]{FFFFFF}WM / BS}                                                                                                         & \multicolumn{1}{l|}{\cellcolor[HTML]{FFFFFF}AG / ES}                                                                                             & \multicolumn{1}{l|}{\cellcolor[HTML]{FFFFFF}Specific}                                                                                       \\ \hline
\multicolumn{7}{c}{\begin{tabular}[c]{@{}c@{}}*WM and BM stand for white-box model-level knowledge and black-box model-level knowledge, respectively, whereas WS, BS, and GS stand for white-box, \\ black-box, and gray-box system-level knowledge. In some cases, multiple attacks were introduced, which is why their model-level knowledge has two values. \\ *AG and AS stand for attack generic and attack specific, while EG and ES stand for error generic and error specific.\end{tabular}}                                                                                                                                                                                                                                                                                                                                                                                                                                                                                                                         
\end{tabular}
}

\end{sidewaystable}

\begin{sidewaystable}
\caption{Paper Mapping (Part 3)}
\label{tab:taxonomy3}

\resizebox{\columnwidth}{!}{%
\begin{tabular}{|c|l|l|}
\hline
\rowcolor[HTML]{DAE8FC} 
\textbf{Ref}      & \multicolumn{1}{c|}{\cellcolor[HTML]{DAE8FC}{\color[HTML]{333333} \textbf{Model-level results}}}                          & \multicolumn{1}{c|}{\cellcolor[HTML]{DAE8FC}{\color[HTML]{333333} \textbf{System-level results (Failure Propagation)}}}                                                                \\ \hline
\rowcolor[HTML]{FFFFFF} 
\textbf{{[}1{]}}  & \begin{tabular}[c]{@{}l@{}}Losing detection \new{due to increased latency} \\ and subsequently tracking of the target object\end{tabular}             & Crashing into another car                                                                                                                                                              \\ \hline
\rowcolor[HTML]{EFEFEF} 
\textbf{{[}2{]}}  & \begin{tabular}[c]{@{}l@{}}Misdetection of objects and Misprediction of the\\ steering wheel angle\end{tabular}           & Crashing to objects, freezing or going off-route                                                                                                                                       \\ \hline
\rowcolor[HTML]{FFFFFF} 
\textbf{{[}3{]}}  & Misdetection of the vehicle in front                                                                                      & Acceleration and crash into the car in front                                                                                                                                           \\ \hline
\rowcolor[HTML]{EFEFEF} 
\textbf{{[}4{]}}  & \begin{tabular}[c]{@{}l@{}}Misdetection of the correct coordinates of the \\ target bounding box\end{tabular}             & Losing path, stability and collision to surrounding obstacles                                                                                                                          \\ \hline
\rowcolor[HTML]{FFFFFF} 
\textbf{{[}5{]}}  & Misprediction of the steering wheel angle                                                                                 & Going off-road                                                                                                                                                                         \\ \hline
\rowcolor[HTML]{EFEFEF} 
\textbf{{[}6{]}}  & \begin{tabular}[c]{@{}l@{}}Misclassification of the adversarial patch as a real \\ human\end{tabular}                     & \begin{tabular}[c]{@{}l@{}}Following the adversarially patched image and failing to detect \\ the real human\end{tabular}                                                              \\ \hline
\rowcolor[HTML]{FFFFFF} 
\textbf{{[}7{]}}  & Failure to detect stop signs and pedestrians                                                                              & Running a stop sign and colliding with pedestrians                                                                                                                                     \\ \hline
\rowcolor[HTML]{EFEFEF} 
\textbf{{[}8{]}}  & \begin{tabular}[c]{@{}l@{}}Detecting fake objects as real while failing to detect \\ a real object\end{tabular}           & Stopping unexpectedly or causing collisions with objects                                                                                                                               \\ \hline
\rowcolor[HTML]{FFFFFF} 
\textbf{{[}9{]}}  & Misclassification of traffic light colors                                                                                 & \begin{tabular}[c]{@{}l@{}}Running a red light, causing a car crash or emergency stop, \\ which leads to the vehicle freezing\end{tabular}                                             \\ \hline
\rowcolor[HTML]{EFEFEF} 
\textbf{{[}10{]}} & Failing to predict the correct reaction to traffic.                                                                       & Causing a collision with the car in front                                                                                                                                              \\ \hline
\rowcolor[HTML]{FFFFFF} 
\textbf{{[}11{]}} & \begin{tabular}[c]{@{}l@{}}Misprediction of the steering wheel angle and failure \\ to detect traffic lights\end{tabular} & \begin{tabular}[c]{@{}l@{}}Losing lane, going offroad, crashing into buildings, ignoring \\ traffic lights, going to crossroads\end{tabular}                                           \\ \hline
\rowcolor[HTML]{EFEFEF} 
\textbf{{[}12{]}} & Misprediction of the steering wheel angle                                                                                 & \begin{tabular}[c]{@{}l@{}}Driving off-road, colliding with road curbs, crashing into \\ obstacles, and having a car crash with oncoming traffic\end{tabular}                          \\ \hline
\rowcolor[HTML]{FFFFFF} 
\textbf{{[}13{]}} & Failing to detect an adversarial object                                                                                   & Crashing into the adversarial object                                                                                                                                                   \\ \hline
\rowcolor[HTML]{EFEFEF} 
\textbf{{[}14{]}} & Detecting the perturbation as a real line                                                                                 & Following the fake line into oncoming traffic                                                                                                                                          \\ \hline
\rowcolor[HTML]{FFFFFF} 
\textbf{{[}15{]}} & Detecting adversarial objects as real cars                                                                                & Stopping completely or suddenly changing lane                                                                                                                                          \\ \hline
\rowcolor[HTML]{EFEFEF} 
\textbf{{[}16{]}} & Misdetection or failure to detect cars or pedestrians                                                                     & Emergency braking or unsafe acceleration leading to a car crash                                                                                                                        \\ \hline
\rowcolor[HTML]{FFFFFF} 
\textbf{{[}17{]}} & Incorrectly detecting lines (wrong control decisions)                                                                     & \begin{tabular}[c]{@{}l@{}}Moving into another lane or veering out of bounds, resulting in \\ collisions with road walls\end{tabular}                                                  \\ \hline
\rowcolor[HTML]{EFEFEF} 
\textbf{{[}18{]}} & Misprediction of the steering wheel angle                                                                                 & Collision with the poster                                                                                                                                                              \\ \hline
\rowcolor[HTML]{FFFFFF} 
\textbf{{[}19{]}} & \begin{tabular}[c]{@{}l@{}}Detecting phantoms as real road signs or real-world \\ objects\end{tabular}                    & \begin{tabular}[c]{@{}l@{}}Acting according to the detected object (Brake or decelerate or follow\\  a phantom lane into collision or into the upcoming lane and, etc...)\end{tabular} \\ \hline
\rowcolor[HTML]{EFEFEF} 
\textbf{{[}20{]}} & Detecting nonexistent obstacles                                                                                           & Emergency braking or the vehicle freezing                                                                                                                                              \\ \hline
\rowcolor[HTML]{FFFFFF} 
\textbf{{[}21{]}} & Misprediction of the steering wheel angle                                                                                 & Moving to the right and then going off the road                                                                                                                                        \\ \hline
\end{tabular}
}

\end{sidewaystable}

\end{document}